\documentclass[10pt, aps, notitlepage]{revtex4-2}
\usepackage{amsmath}
\usepackage{amssymb}

\usepackage{bbold} %to support the mathematical symbol for identiy.
\usepackage{braket}

\usepackage{graphicx}% Include figure files
\graphicspath{{figures}}
\usepackage{subcaption} %Support for subfigures.
\usepackage{dcolumn}% Align table columns on decimal point
\usepackage{bm}% bold math
\usepackage{hyperref}% add hypertext capabilities
\usepackage[normalem]{ulem}
\hypersetup{
    colorlinks=true,
    linkcolor=blue,
    filecolor=magenta,      
    urlcolor=cyan,
    }

\begin{document}

\title{Polarization and correlation effects in $\gamma\gamma \to W^-W^+$ at Next-to-Leading-Order
Electroweak Accuracy}

\author{Jun Jiang}
\email{jiangjun87@sdu.edu.cn}
\affiliation{School of Physics, Shandong University, Jinan, Shandong 250100, China}

\author{Peng-Cheng Lu}
\email{pclu@sdu.edu.cn}
\affiliation{School of Physics, Shandong University, Jinan, Shandong 250100, China}

\author{Zongguo Si}
\email{zgs@sdu.edu.cn}
\affiliation{School of Physics, Shandong University, Jinan, Shandong 250100, China}

\author{Han Zhang}
\email{han.zhang@mail.sdu.edu.cn}
\affiliation{School of Physics, Shandong University, Jinan, Shandong 250100, China}

\author{Xin-Yi Zhang}
\email{sps\_zhangxy@ujn.edu.cn}
\affiliation{School of Physics and Technology, University of Jinan, Jinan, Shandong 250022, China}

\date{\today}

\begin{abstract}
    We study the next-to-leading order (NLO) electroweak (EW) corrections to the polarization and correlation of $W$ boson pairs produced through photon-photon fusion in proton-proton ultra-peripheral collisions at the LHC, with both $W$ bosons decaying leptonically. The azimuthal correlation distributions of the charged leptons, which receive no contribution from anomalous quartic gauge couplings (AQGCs) at linear order, are perturbatively stable at the 0.40\% level, whereas individual correlation coefficient $C_{2,2}$ receives relative corrections of up to $-33.84\%$. Some of correlation coefficients that vanish at leading order (LO) in the Standard Model but are sensitive to AQGCs acquire non-zero one-loop values of order $10^{-4}-10^{-3}$. We quantify the region of the ($a_0/\Lambda^2,a_c/\Lambda^2$) plane in which these loop-induced baselines compete with the AQGC contributions and find that, while irrelevant at the current experimental sensitivity, they must be included once the correlation coefficients are measured at the per-mille level.
\end{abstract}

\maketitle
\section{Introduction}

Although photon-photon ($\gamma\gamma$) induced vector boson scattering (VBS)
provides a direct probe of both the Standard Model (SM) and anomalous
quartic gauge couplings (AQGCs), its measurement in proton-proton ($pp$)
collisions is intrinsically challenging due to the heavy dominance of quark
parton distribution functions (PDFs) over the photon PDF.
However, the
ultra-peripheral collision (UPC) mode of the LHC, characterized by minimal
hadronic background, renders the study of $\gamma\gamma$ induced VBS not
only feasible, but a vital complement to traditional hadronic scattering
measurements. In this context, $W^- W^+$ pair production via $\gamma\gamma$
annihilation has drawn significant interest. Initial measurements of this
process in $p \bar{p}$ collisions at Tevatron were performed by the D0 Collaboration~\cite{D0:2013rce},
with the resulting constraints on AQGCs rapidly superseded by
CMS~\cite{CMS:2013hdf} using $pp$ UPCs at LHC. Over the past decade, both the
CMS~\cite{CMS:2013hdf, CMS:2016rtz, CMS:2022dmc} and
ATLAS~\cite{ATLAS:2016lse, ATLAS:2016snd, ATLAS:2020iwi} collaborations
have progressively refined these measurements using higher center of mass (CM)
energies and increased integrated luminosity. The first observation of
$\gamma\gamma \to W^- W^+$ in $pp$ UPCs was established by ATLAS with a
significance of $8.4\,\sigma$~\cite{ATLAS:2020iwi}, while the most stringent bounds
on dimension-eight AQGCs to date have been reported by CMS~\cite{CMS:2026job}.

On the theoretical front, NLO electroweak (EW) corrections to polarized $W^-W^+$
production via $\gamma\gamma$ annihilation were calculated in
Refs.~\cite{Denner:1995jv,Denner:1996wm}, with the dependence on the
initial photon polarizations subsequently investigated in
Ref.~\cite{Jikia:1996uu}. More recently, automated NLO EW predictions
for the unpolarized $\gamma\gamma\to W^-W^+$ cross section in various
UPC configurations have become available
in Ref.~\cite{Shao:2025bma}. Early studies of the sensitivity of this
process to AQGCs at the LHC
\cite{Belanger:1992qi,Pierzchala:2008xc,Chapon:2009hh} provided important
theoretical input for the experimental searches discussed above.
For quark-initiated $W$-pair production at the LHC, NLO QCD and EW
corrections to polarized cross sections have also been presented in
Refs.~\cite{Denner:2020bcz,Denner:2023ehn}.

In light of the anticipated experimental precision in upcoming LHC runs, a
comprehensive determination of the polarization and correlation effects
of the final state $W$ boson pair is both timely and essential.
This direction remains largely unexplored experimentally, and theoretical
predictions incorporating NLO EW corrections for these polarization and
correlation observables are currently lacking. The goal of this work is to fill
this gap by systematically investigating the impact of NLO EW corrections on the
$W$-pair polarization and correlation structure. A related recent study
demonstrated that correlation asymmetries offer enhanced sensitivity to
anomalous $WWWW$ interactions in same sign $W$ boson
scattering~\cite{Eboli:2026mbq}.

Polarization and correlations are naturally formulated within the density
matrix framework. In this paper, we expand the density matrix in an orthogonal basis of $3\times3$ Hermitian matrices which, in contrast to the Gell-Mann basis employed in Ref.~\cite{Ashby-Pickering:2022umy}, is adapted to the angular structure of the $W$ decay. This basis consists of the three Cartesian spin operators and five rank-two tensor operators, and its decay angular coefficients are mutually orthogonal. From this parametrization, we derive a set of dual angular functions that enable the experimental extraction of all polarization and correlation coefficients by moment projection.
Experimentally, measuring these angular distributions requires full kinematic
reconstruction of the process, which is a formidable task given the presence of two
invisible neutrinos in the leptonic decay modes.
Forward proton tagging, available at the LHC via the CMS Precision
Proton Spectrometer (PPS)~\cite{Bossini:2020ycc} and the ATLAS
Forward Proton (AFP) detector~\cite{Trzebinski:2023lzg, Lysenko:2026hql},
enables direct determination of the initial state photon momenta.
When combined with central detector measurements, this information
significantly improves event reconstruction for polarization and
correlation studies in $pp$ UPCs.

This paper is organized as follows. In Section~\ref{sec:DensityMatrix}, we
review the density matrix formalism, present our proposed parametrization, and
briefly outline the effective field theory Lagrangian for AQGCs. Section~\ref{sec:Numerics} presents and discusses
our numerical results. Finally, Section~\ref{sec:Summary} provides our
concluding remarks.

\section{Theoretical Framework}
\label{sec:DensityMatrix}
We consider $\gamma\gamma$ annihilation into a pair of oppositely charged $W$ bosons
that decay fully leptonically. At LO, it's described by
\begin{equation}
    \gamma\gamma \to W^-W^+ \to
    l_1^-\bar{\nu}_{l_1} l_2^+\nu_{l_2},
    \label{eq:full-process}
\end{equation}
where $l_1$ and $l_2$ denote final-state leptons of identical or different flavors.
In this work, we employ the narrow width approximation (NWA), which retains only
contributions with two resonant $W$ bosons and neglects single resonant,
non-resonant, and non-factorizable contributions. Non-factorizable corrections
can be consistently incorporated within the double pole approximation (DPA)~\cite{Bredenstein:2005zk}.
An extension of the density matrix formalism adopted here to the DPA is left
for future work. Within the NWA, the full process in
Eq.~(\ref{eq:full-process}) factorizes into distinct production and decay
subprocesses, which can, in principle, be computed independently to arbitrary
precision.

To resolve the polarization and correlation structure within the NWA, the
polarization states of the intermediate $W$ bosons must be specified explicitly.
In their rest frame, these are defined by the polarization vectors
\begin{equation}
    \begin{split}
    &\epsilon_{R} = \frac{1}{\sqrt{2}}(0, -1, -i, 0),\\
    &\epsilon_{0} = (0, 0, 0, 1),\\
    &\epsilon_{L} = \frac{1}{\sqrt{2}}(0, 1, -i, 0),
    \end{split}
    \label{eq:polr}
\end{equation}
corresponding to right-handed ($R$), longitudinal ($0$), and left-handed ($L$)
polarizations, respectively. To render Eq.~(\ref{eq:polr}) fully unambiguous, one
must first specify the three Cartesian axes ($x$, $y$, $z$) in the rest frame of the corresponding
$W$ boson. We define these axes explicitly below.

\subsection{Production density matrix}
We first address the production subprocess
\begin{equation}
    \gamma\left(p_1, \lambda_1\right) + \gamma\left(p_2, \lambda_2\right) \to 
    W^-\left(k_1, \lambda^-\right) + W^+\left(k_2, \lambda^+\right),
    \label{eq:process}
\end{equation}
where $p_1, p_2, k_1$, and $k_2$ represent the four momenta of the respective particles in the 
$W^-W^+$ CM frame, and $\lambda_1, \lambda_2, \lambda^-$, and $\lambda^+$ 
denote their polarization states in this frame. The corresponding scattering amplitude is given by
\begin{equation}
    \mathcal{M}^{\lambda_1, \lambda_2, \lambda^-, \lambda^+} =
    \bra{W^-\left(k_1, \lambda^-\right) W^+\left(k_2, \lambda^+\right)} \mathcal{M}
    \ket{\gamma\left(p_1, \lambda_1\right)\gamma\left(p_2, \lambda_2\right)}.
    \label{eq:amplitude22}
\end{equation}

We now turn to defining the polarization vectors for moving $W$ bosons. For the $W^-$ boson 
carrying momentum $k_1 = \left(E_{k_1}, \mathbf{k}_1\right)$, the longitudinal polarization vector 
is defined as
\begin{equation}
    \epsilon_{0}\left(k_1\right) = \left(\frac{|\mathbf{k}_1|}{m_W}, \frac{E_{k_1}}{m_W}\hat{\mathbf{k}}_1\right),
\end{equation}
which reduces to $\epsilon_0$ in Eq.~(\ref{eq:polr}) when boosted along $\mathbf{k}_1$ to 
the rest frame of $W^-$, establishing the local $z$ axis. Here $m_W$ is the mass of $W$ boson, $\mathbf{k}_1$ is the spatial momentum of $W^-$, 
$\hat{\mathbf{k}}_1$ is its unit vector representing the direction of $\mathbf{k}_1$, and $E_{k_1} = \sqrt{\mathbf{k}_1^2 + m_W^2}$ is the 
energy of $W^-$. Constructing the transverse polarization vectors $\epsilon_L\left(k_1\right)$ 
and $\epsilon_R\left(k_1\right)$ requires specifying local $x$ and $y$ directions. Here, 
we adopt the standard orthonormal basis $\left\{\hat{\mathbf{n}}_1, \hat{\mathbf{r}}_1, 
\hat{\mathbf{k}}_1\right\}$. Let $\hat{\mathbf{p}}_1$ denote the direction of the incoming photon with momentum 
$p_1$ in the $W^-W^+$ CM frame, then the transverse unit vectors $\hat{\mathbf{n}}_1$ and $\hat{\mathbf{r}}_1$ are defined by
\begin{equation}
    \hat{\mathbf{n}}_1 = \frac{\hat{\mathbf{p}}_1\times \hat{\mathbf{k}}_1}
    {|\hat{\mathbf{p}}_1\times \hat{\mathbf{k}}_1|}\ \text{and}\ 
    \hat{\mathbf{r}}_1 = \hat{\mathbf{k}}_1\times \hat{\mathbf{n}}_1.
    \label{eq:hatnr1}
\end{equation}
Because $\hat{\mathbf{n}}_1$ and $\hat{\mathbf{r}}_1$ 
are invariant under boost along $\mathbf{k}_1$, they directly fix the local $x$ and $y$ 
axes in the $W^-$ rest frame, thereby uniquely identifying $\epsilon_L$ and $\epsilon_R$ in 
Eq.~(\ref{eq:polr}). The corresponding vectors $\epsilon_L\left(k_1\right)$ and 
$\epsilon_R\left(k_1\right)$ are obtained via a Lorentz boost back to $W^-W^+$ CM frame along the 
$z$ axis. An analogous construction defines the unit vectors $\hat{\mathbf{n}}_2$ and 
$\hat{\mathbf{r}}_2$ for the $W^+$ boson:
\begin{equation}
    \hat{\mathbf{n}}_2 = \frac{\hat{\mathbf{p}}_2\times \hat{\mathbf{k}}_2}
    {|\hat{\mathbf{p}}_2\times \hat{\mathbf{k}}_2|}\ \text{and}\ 
    \hat{\mathbf{r}}_2 = \hat{\mathbf{k}}_2\times \hat{\mathbf{n}}_2,
    \label{eq:hatnr2}
\end{equation}
where the intermediate steps follow identically and are omitted for brevity. Then the polarization vectors of $W^+$
are also specified without ambiguity.

In perturbation theory, $\mathcal{M}^{\lambda_1, \lambda_2, \lambda^-, \lambda^+}$ in Eq.~(\ref{eq:amplitude22}) admits 
a power series expansion in the electric charge $e$. Denoting the $\mathcal{O}\left(e^2\right)$ 
LO term by $\mathcal{M}^{\lambda_1, \lambda_2, \lambda^-, \lambda^+}_{LO}$ and 
the $\mathcal{O}\left(e^4\right)$ one-loop virtual correction by 
$\mathcal{M}^{\lambda_1, \lambda_2, \lambda^-, \lambda^+}_{V}$, the LO
contribution to the production density matrix takes the form
\begin{equation}
    \rho_{LO}^{\lambda^-_{1}, \lambda^-_{2}, \lambda^+_{1}, \lambda^+_{2}} =
    \sum\limits_{\lambda_1, \lambda_2}
    \mathcal{M}_{LO}^{\lambda^-_{1}, \lambda^+_{1}, \lambda_{1}, \lambda_{2}}
    (\mathcal{M}_{LO}^{\lambda^-_{2}, \lambda^+_{2}, \lambda_{1}, \lambda_{2}})^{*},
    \label{eq:ProdDenMatrixLO}
\end{equation}
which is of order $\mathcal{O}\left(\alpha^2\right)$, with $\alpha = \frac{e^2}{4\pi}$ 
denoting the fine-structure constant. The conventional polarization summed squared 
amplitude at LO is recovered via
\begin{equation}
    |\mathcal{M}_{LO}|^2 = \operatorname{Tr}\left[\rho_{LO}\right].
\end{equation}

The virtual correction to the production density matrix, arising from the interference 
between the LO and one-loop amplitudes, is given by
\begin{equation}
    \rho_{V}^{\lambda^-_{1}, \lambda^-_{2}, \lambda^+_{1}, \lambda^+_{2}} =
    \sum\limits_{\lambda_1, \lambda_2}\left(
    \mathcal{M}_{LO}^{\lambda^-_{1}, \lambda^+_{1}, \lambda_{1}, \lambda_{2}}
    (\mathcal{M}_{V}^{\lambda^-_{2}, \lambda^+_{2}, \lambda_{1}, \lambda_{2}})^{*} +
    \mathcal{M}_{V}^{\lambda^-_{1}, \lambda^+_{1}, \lambda_{1}, \lambda_{2}}
    (\mathcal{M}_{LO}^{\lambda^-_{2}, \lambda^+_{2}, \lambda_{1}, \lambda_{2}})^{*}
    \right),
    \label{eq:ProdDenMatrixV}
\end{equation}
which is of order $\mathcal{O}\left(\alpha^3\right)$. Similarly, the corresponding 
$\mathcal{O}\left(\alpha^3\right)$ correction to the polarization summed squared 
amplitude is obtained from
\begin{equation}
    2\operatorname{Re}\left[\mathcal{M}^*_{LO}\mathcal{M}_{V}\right] = 
    \operatorname{Tr}\left[\rho_{V}\right].
    \label{eq:TrRhoV}
\end{equation}

At $\mathcal{O}\left(\alpha^3\right)$, the real emission process
\begin{equation}
    \gamma(p_{1}, \lambda_1) + \gamma(p_{2}, \lambda_2) \to 
    W^{-}(k_{1}, \lambda^-) + W^{+}(k_{2}, \lambda^+) + \gamma\left(k_{3}, \lambda_3\right)
    \label{eq:real-emission}
\end{equation}
must also be included, where $k_3$ and $\lambda_3$ specify the momentum and polarization 
state of the radiated photon in the $W^-W^+$ CM frame. Since our primary interest lies 
in the polarizations and correlations of the two $W$ bosons, the real emission 
contribution to the production density matrix is obtained by summing over the final state 
photon polarizations:
\begin{equation}
    \rho_{R}^{\lambda^-_{1}, \lambda^-_{2}, \lambda^+_{1}, \lambda^+_{2}} =
    \sum\limits_{\lambda_1, \lambda_2, \lambda_3}
    \mathcal{M}_{R}^{\lambda^-_{1}, \lambda^+_{1}, \lambda_{3}, \lambda_{1}, \lambda_{2}}
    (\mathcal{M}_{R}^{\lambda^-_{2}, \lambda^+_{2}, \lambda_{3}, \lambda_{1}, \lambda_{2}})^{*},
    \label{eq:ProdDenMatrixReal}
\end{equation}
with
\begin{equation}
    \mathcal{M}_{R}^{\lambda^-, \lambda^+, \lambda_3, \lambda_1, \lambda_2} =
    \bra{W^{-}\left(k_{1}, \lambda^-\right), W^{+}\left(k_{2}, \lambda^+\right), 
    \gamma\left(k_{3}, \lambda_3\right)}\mathcal{M}
    \ket{\gamma\left(p_{1}, \lambda_1\right), \gamma\left(p_{2}, \lambda_2\right)},
\end{equation}
retaining only the leading $\mathcal{O}\left(e^3\right)$ terms in 
$\mathcal{M}_{R}^{\lambda^-, \lambda^+, \lambda_3, \lambda_1, \lambda_2}$. Again, the 
polarization summed squared amplitude for the real emission process is given by
\begin{equation}
    |\mathcal{M}_{R}|^2 = \operatorname{Tr}\left[\rho_{R}\right].
    \label{eq:SquaredAmpReal}
\end{equation}

We next formulate the expansion of the production density matrix. Being 
$(3\times 3)\otimes(3\times 3)$ Hermitian matrices, $\rho_{LO}$, $\rho_V$, and $\rho_R$ 
can each be expanded in terms of a basis of traceless Hermitian matrices supplemented by one identity matrix. Although the 
eight Gell-Mann matrices form a standard choice~\cite{Ashby-Pickering:2022umy}, we propose 
parametrizing the density matrix using the following set of eight traceless Hermitian 
matrices for reasons that will be explained later:
\begin{equation}
    \begin{split}
        R^1 &= \frac{\sqrt{3}}{\sqrt{2}}\left(
        \begin{array}{ccc}
         1 & 0 & 0 \\
         0 & 0 & 0 \\
         0 & 0 & -1 \\
        \end{array}
        \right),\quad
        R^2 = \frac{\sqrt{3}}{\sqrt{2}}\left(
        \begin{array}{ccc}
         0 & 0 & -i \\
         0 & 0 & 0 \\
         i & 0 & 0 \\
        \end{array}
        \right),\quad
        R^3 = \frac{\sqrt{3}}{\sqrt{2}}\left(
        \begin{array}{ccc}
         0 & 0 & -1 \\
         0 & 0 & 0 \\
         -1 & 0 & 0 \\
        \end{array}
        \right),\quad\\
        R^4 &= \frac{\sqrt{3}}{2}\left(
        \begin{array}{ccc}
         0 & 1 & 0 \\
         1 & 0 & -1 \\
         0 & -1 & 0 \\
        \end{array}
        \right),\quad
        R^5 = \frac{\sqrt{3}}{2}\left(
        \begin{array}{ccc}
         0 & -i & 0 \\
         i & 0 & -i \\
         0 & i & 0 \\
        \end{array}
        \right),\quad
        R^6 = \frac{\sqrt{3}}{2}\left(
        \begin{array}{ccc}
         0 & 1 & 0 \\
         1 & 0 & 1 \\
         0 & 1 & 0 \\
        \end{array}
        \right),\\
        R^7 &= \frac{\sqrt{3}}{2}\left(
        \begin{array}{ccc}
         0 & -i & 0 \\
         i & 0 & i \\
         0 & -i & 0 \\
        \end{array}
        \right),\quad
        R^8 = \frac{1}{\sqrt{2}}\left(
        \begin{array}{ccc}
         1 & 0 & 0 \\
         0 & -2 & 0 \\
         0 & 0 & 1 \\
        \end{array}
        \right).\\
    \end{split}
    \label{eq:Definition-of-Ri}
\end{equation}
The signs and ordering of the $R^i$ matrices are chosen such that $\frac{\sqrt{2}}{\sqrt{3}}R^i$ 
satisfy an algebra analogous to the Gell-Mann matrices, spanning the Lie algebra of $SU(3)$.
Crucially, three of these matrices relate directly to the spin 
operators of the $W^-$ boson via
\begin{equation} 
    R^1 = \frac{\sqrt{3}}{\sqrt{2}}S_z,\quad
    R^5 = \frac{\sqrt{3}}{\sqrt{2}}S_y,\quad
    R^6 = \frac{\sqrt{3}}{\sqrt{2}}S_x,
\end{equation}
where $S_i$ ($i = x, y, z$) represents the Cartesian spin operators of $W^-$.
Moreover,
\begin{equation}
    \begin{split}
        R^2 &= \frac{\sqrt{3}}{\sqrt{2}}\left\{S_x, S_y\right\},\quad
        R^3 = -\frac{\sqrt{3}}{\sqrt{2}}\left(S_x^2 - S_y^2\right),\quad
        R^4 = \frac{\sqrt{3}}{\sqrt{2}}\left\{S_x, S_z\right\},\quad\\
        R^7 &= \frac{\sqrt{3}}{\sqrt{2}}\left\{S_y, S_z\right\},\quad
        R^8 = \frac{1}{\sqrt{2}}\left(3 S_z^2 - 2 \mathbb{1}\right),
    \end{split}
\end{equation}
where $\left\{A, B\right\} = AB + BA$ denotes the anticommutator
of $A$ and $B$.

The density matrix expansion then reads
\begin{equation}
    \rho_{L} = \frac{1}{9} \left(
    A^{L}\ \mathbb{1}\otimes\mathbb{1} +
    \sum_{i = 1}^{8} B^{L}_i\ R^i\otimes\mathbb{1} + 
    \sum_{i = 1}^{8} \bar{B}^{L}_i\ \mathbb{1}\otimes R^i + 
    \sum_{i,j = 1}^{8} C_{i,j}^{L}\ R^i\otimes R^j
    \right),
    \label{eq:Param-of-RhoL}
\end{equation}
where $L \in \{LO, V, R\}$. The scalar $A^{L} = \operatorname{Tr}\left[\rho_{L}\right]$ gives 
the corresponding polarization summed squared amplitude, while
\begin{equation}
    B_i^L = \operatorname{Tr}\left[\rho_L\cdot\left(R^i\otimes\mathbb{1}\right)\right]\
    \mathrm{and}\
    \bar{B}_i^L = \operatorname{Tr}\left[\rho_L\cdot\left(\mathbb{1}\otimes R^i\right)\right]
\end{equation}
denote polarization coefficients, and
\begin{equation}
    C_{i,j}^L = \operatorname{Tr}\left[\rho_L\cdot\left(R^i\otimes R^j\right)\right]
\end{equation}
represent the correlation coefficients. The LO contribution 
$A^{LO}$ evaluates to
\begin{equation}
    A^{LO} = \operatorname{Tr}\left[\rho_{LO}\right] = |\mathcal{M}_{LO}|^2 =
    \frac{128 \pi ^2 \alpha ^2 \left(3 \beta ^4 \left(y^4-2 y^2+2\right)+
    2 \beta ^2 \left(8 y^2-3\right)+19\right)}{\left(\beta ^2 y^2-1\right)^2},
\end{equation}
where $\beta = \sqrt{1 - \frac{4m_W^2}{s}}$ is the $W$ boson velocity and $y = \cos\vartheta$ is the 
cosine of the scattering angle between the incoming photon with momentum $p_1$ and the outgoing $W^-$ boson in 
$W^-W^+$ CM frame. Analytical expressions for the LO coefficients $B^{LO}_i$, $\bar{B}^{LO}_i$, and 
$C^{LO}_{i,j}$ are listed in Appendix~\ref{app:MCE}. Because the full NLO EW expressions are 
extraordinarily unwieldy, we evaluate them numerically in this paper.

\subsection{Anomalous Quartic Gauge Couplings}

Apart from SM contributions, we also consider 
contributions from AQGCs defined 
through the effective Lagrangian~\cite{Belanger:1992qi, Pierzchala:2008xc, 
Chapon:2009hh}
\begin{equation}
    \mathcal{L}_{6} = - \frac{e^2 a_0}{8 \Lambda^2}F_{\mu\nu}F^{\mu\nu}
    W^{+\alpha}W^-_{\alpha} - \frac{e^2 a_c}{16 \Lambda^2} F_{\mu\alpha}
    F^{\mu\beta}(W^{+\alpha}W^-_\beta + W^{-\alpha}W^+_\beta),
    \label{eq:effective-lagrangian}
\end{equation}
where $F_{\mu\nu} = \partial_{\mu}A_{\nu} - \partial_{\nu}A_{\mu}$ is the 
photon field strength tensor, and $W^+$ ($W^-$) denotes the field corresponding 
to the positively (negatively) charged $W$ boson. The dimensionless parameters 
$a_{0}$ and $a_{c}$ represent neutral and charged AQGCs, respectively, while 
$\Lambda$ specifies the characteristic high energy scale for the AQGCs. The 
subscript $6$ in $\mathcal{L}_6$ indicates that the operators in this 
effective Lagrangian are of dimension six.

In this work, we neglect the interference between the one-loop SM amplitude 
and the anomalous amplitude. Moreover, we keep only terms that are linear in $a_0$ 
and $a_c$. To prevent violation of unitarity, we introduce 
dipole form factors as in Refs.~\cite{Pierzchala:2008xc, Chapon:2009hh} to 
regulate the rapid growth of the $\gamma\gamma\to W^+W^-$ cross section at 
energy scales accessible by current experimental facilities:
\begin{equation}
    a_{0, c} \to \frac{a_{0, c}}
    {\left(1 + \frac{s}{\Lambda^{2}_{\text{cutoff}}}\right)^2},
\end{equation}
where $s = (p_1 + p_2)^2$, $\Lambda_{\text{cutoff}}$ is the cutoff scale, commonly chosen as 
$\Lambda_{\text{cutoff}} = 500\ \text{GeV}$ in experimental 
searches~\cite{CMS:2016rtz}.

The contribution to the production density matrix from AQGCs is then
given by
\begin{equation}
    \rho_{NP}^{\lambda^-_{1}, \lambda^-_{2}, \lambda^+_{1}, \lambda^+_{2}} =
    \sum\limits_{\lambda_1, \lambda_2}\left(
    \mathcal{M}_{LO}^{\lambda^-_{1}, \lambda^+_{1}, \lambda_{1}, \lambda_{2}}
    (\mathcal{M}_{NP}^{\lambda^-_{2}, \lambda^+_{2}, \lambda_{1}, \lambda_{2}})^{*} +
    \mathcal{M}_{NP}^{\lambda^-_{1}, \lambda^+_{1}, \lambda_{1}, \lambda_{2}}
    (\mathcal{M}_{LO}^{\lambda^-_{2}, \lambda^+_{2}, \lambda_{1}, \lambda_{2}})^{*}
    \right),
    \label{eq:ProdDenMatrixNP}
\end{equation}
where the subscript ``NP'' denotes ``New Physics'', and 
$\mathcal{M}_{NP}^{\lambda^-, \lambda^+, \lambda_{1}, \lambda_{2}}$ is 
the tree level contribution to the scattering amplitude for the process in 
Eq.~(\ref{eq:process}) derived from $\mathcal{L}_6$ in 
Eq.~(\ref{eq:effective-lagrangian}).

We decompose $\rho_{NP}$ 
following the same parametrization as in Eq.~(\ref{eq:Param-of-RhoL}) with 
$L = NP$.
We find that
\begin{equation}
    A^{NP} = \operatorname{Tr}\left[\rho_{NP}\right] = 0.
\end{equation}
Therefore, at linear order in $a_0$ and $a_c$, the AQGCs in Eq.~(\ref{eq:effective-lagrangian}) cannot be probed 
through the unpolarized kinematic distributions of the production process. The quadratic terms are suppressed by an additional power of $a_{0,c}/\Lambda^2$.
Polarization and correlation effects of the two resonant $W$ bosons are 
consequently the natural observables for such searches at linear order. Since no CP violation is introduced in 
Eq.~(\ref{eq:effective-lagrangian}), we have $B^{NP}_i = \bar{B}^{NP}_i$ and $C^{NP}_{j,i} = C^{NP}_{i,j}$. 
Explicit expressions for $B^{NP}_i$ and $C^{NP}_{i,j}$ are provided in 
Appendix~\ref{app:MCE}.
An important feature of the polarization and correlation coefficients presented in
Appendix~\ref{app:MCE} is the clean separation of the SM LO and AQGC contributions.
In particular, for coefficients that vanish at LO in the SM, the
experimental sensitivity to
$a_0$ and $a_c$ is limited by the
SM loop-induced contributions, which we quantify in Sec.~\ref{sec:Numerics}.

Combining the LO SM contribution, the NLO EW corrections, and the AQGC contributions,
the total cross section $\hat{\sigma}$ for $W$ pair production in $\gamma\gamma$ collisions at a fixed $\gamma\gamma$ CM energy is given by
\begin{equation}
    \hat{\sigma} = \frac{1}{2 s}\sum_{L \in \left\{LO, V, R, NP\right\}} \int \frac{1}{4} A^{L} d\Pi_L,
    \label{eq:parton-cross-section}
\end{equation}
where $d\Pi_{LO} = d\Pi_V = d\Pi_{NP} = d\Pi_2$, and $d\Pi_R = d\Pi_3$,
with $d\Pi_2$, $d\Pi_3$ denoting the Lorentz invariant 2- and 3-body 
phase space elements for the processes in Eqs.~(\ref{eq:process}) and (\ref{eq:real-emission}), respectively.

Phase space integrated polarization and correlation coefficients are denoted by $\hat{B}_i$, 
$\hat{\bar{B}}_i$, and $\hat{C}_{i,j}$. Specifically, for the polarization coefficients $\hat{B}_i$, 
we have
\begin{equation}
    \hat{B}_i = \frac{1}{\hat{\sigma}} \frac{1}{2 s}
    \sum_{L \in \left\{LO, V, R, NP\right\}} \int \frac{1}{4} B_i^{L}d\Pi_L.
    \label{eq:parton-pol}
\end{equation}
Analogous definitions hold for $\hat{\bar{B}}_i$ and $\hat{C}_{i,j}$. The caret notation 
($\hat{\ }$) explicitly indicates observables evaluated at fixed $\gamma\gamma$ CM 
energy. In realistic collider scenarios where photons are emitted from protons, heavy ions, 
or charged leptons, one must introduce additional integrations over the appropriate photon 
PDFs. The resulting observables are denoted by the 
same symbols with the carets removed.

\subsection{Decay Density Matrix}

For the leptonic decay process
\begin{equation}
    W^-(\lambda^-) \to l^-(q_1, s_1) + \bar{\nu}_l(q_2, s_2),
\end{equation}
the decay density matrix, obtained by summing over final state lepton helicities, is defined as
\begin{equation}
    \rho_{-}^{\lambda^-_1,\lambda^-_2}=
    \sum\limits_{s_1, s_2}\mathcal{M}_{\mathrm{decay}}^{\lambda^-_2, s_1, s_2}
    (\mathcal{M}_{\mathrm{decay}}^{\lambda^-_1, s_1, s_2})^*.
\end{equation}
Here,
\begin{equation}
    \mathcal{M}_{\mathrm{decay}}^{\lambda^-, s_1, s_2} =
    \bra{l^-(q_1, s_1), \bar{\nu}_l(q_2, s_2)}\mathcal{M}_{\mathrm{decay}}
    \ket{W^-(\lambda^-)}
\end{equation}
denotes the helicity amplitude for $W^-$ decay in its rest frame, where the superscript
$\lambda^-$ specifies the $W^-$ polarization state, $s_1$ and $s_2$ denote the helicities of
$l^-$ and $\bar{\nu}_l$, and $q_1$ and $q_2$ represent their respective four momenta.
For convenience, we introduce the normalized decay density matrix
\begin{equation}
    \tilde{\rho}_{-} = \frac{\rho_{-}}{\operatorname{Tr}\left[\rho_{-}\right]}.
\end{equation}
The polarization summed squared amplitude for the decay process is given by
\begin{equation}
    |\mathcal{M}_{\mathrm{decay}}|^2 = \operatorname{Tr}\left[\rho_{-}\right].
\end{equation}

At LO in the SM, neglecting the charged lepton mass $m_l$ and
treating $\bar{\nu}_l$ as massless, a direct evaluation yields
\begin{equation}
    \tilde{\rho}_{-}(\theta, \phi)=
\left(
\begin{array}{ccc}
 \cos ^4\frac{\theta }{2} & \frac{e^{-i \phi } \sin \theta  \left(1 + \cos \theta\right)}{2 \sqrt{2}} & \frac{1}{4} e^{-2 i \phi } \sin ^2\theta\\
    \frac{e^{i \phi } \sin \theta\left(1 + \cos \theta \right)}{2 \sqrt{2}} & \frac{\sin ^2\theta}{2} & \frac{e^{-i \phi } \sin \theta \left(1 - \cos \theta\right)}{2\sqrt{2}} \\
 \frac{1}{4} e^{2 i \phi } \sin ^2\theta & \frac{e^{i \phi } \sin \theta \left(1 - \cos \theta\right)}{2\sqrt{2}} & \sin ^4\frac{\theta }{2}\\
\end{array}
\right),
    \label{eq:Rho-M}
\end{equation}
where $\theta$ and $\phi$ denote the polar and azimuthal angles of $l^-$ relative to the
local $z$ axis in the $W^-$ rest frame. Hereafter, the explicit angular dependence of
$\rho_{-}(\theta, \phi)$ and $\tilde{\rho}_{-}(\theta, \phi)$ will be suppressed for brevity.
We note that Eq.~(\ref{eq:Rho-M}) coincides with Eq.~(4.5) of Ref.~\cite{Ashby-Pickering:2022umy}.

Decomposing $\tilde{\rho}_{-}$ in terms of the Hermitian basis matrices $R^i$ defined in
Eq.~(\ref{eq:Definition-of-Ri}), we write
\begin{equation}
    \tilde{\rho}_{-} = \frac{1}{3}\left(\mathbb{1} + \sum_{i = 1}^{8} g_{i}(\theta, \phi)R^i\right),
    \label{eq:Param-of-RhoM}
\end{equation}
where the expansion coefficients are given by $g_{i}(\theta, \phi) =
\operatorname{Tr}[R^i\cdot\tilde{\rho}_{-}]$. Crucially, the coefficients $g_i(\theta, \phi)$ form a
mutually orthogonal set under phase space integration over the decay angular kinematics, a property
not satisfied by the standard Gell-Mann matrices. This orthogonality motivates our choice of the
$R^i$ basis.

It is useful to comment on how
radiative corrections may modify the decay density matrix. For massless leptons,
the virtual
correction to the decay subprocess enters only through an overall constant
multiplicative factor, as can be seen from Eq.~(3.5) of
Ref.~\cite{Bredenstein:2005zk}. Real photon emission does modify the angular distribution of the charged lepton. However, on the basis of rotational covariance, the radiative
corrections can be absorbed into two analyzing powers, $\kappa_V$ and $\kappa_T$,
associated with the vector and tensor components, respectively. Eq.~(\ref{eq:Param-of-RhoM}) would then generalize to
\begin{equation}
    \tilde{\rho}^{NLO}_{-} = \frac{1}{3}\left(\mathbb{1} +
    \kappa_V \sum_{i\in\left\{1, 5, 6\right\}} g_{i}(\theta, \phi)R^i +
    \kappa_T \sum_{i\in\left\{2, 3, 4, 7, 8\right\}} g_{i}(\theta, \phi)R^i
    \right).
\end{equation}
A complete NLO EW treatment of the decay density matrix, including the
determination of $\kappa_V$ and $\kappa_T$, is left for future work.

To facilitate the experimental reconstruction of the production density matrix using the
angular distribution of the final state charged leptons, we define the corresponding
dual angular functions $f_{i}(\theta, \phi)$ through normalization:
\begin{equation}
    f_{i}(\theta, \phi) = 4\pi\frac{g_{i}(\theta, \phi)}
    {\int g_{i}^{2}(\theta, \phi)\ d\Omega_{l^-}},
\end{equation}
where $d\Omega_{l^-} = \sin\theta d\theta d\phi$ is the solid angle element of $l^-$.
These functions satisfy the orthogonality relations
\begin{equation}
    \int f_i g_j\ d\Omega_{l^-} = 4\pi\delta_{i,j},\quad \text{and}\quad
    \int f_i\ d\Omega_{l^-} = 0,
    \label{eq:orthogonality}
\end{equation}
where $\delta_{i,j}$ denotes the Kronecker delta.

The explicit expressions for $f_{i}(\theta, \phi)$ are:
\begin{equation}
    \begin{split}
        f_1 &= \sqrt{6} \cos (\theta ),\quad
        f_2 = 5\sqrt{\frac{3}{2}}\sin ^2(\theta )\sin(2\phi),\quad
        f_3 = -5\sqrt{\frac{3}{2}}\sin ^2(\theta )\cos(2\phi),\\
        f_4 &= 5\sqrt{\frac{3}{2}}\sin(2\theta)\cos(\phi),\quad
        f_5 = \sqrt{6} \sin(\theta )\sin(\phi),\quad
        f_6 = \sqrt{6} \sin(\theta )\cos(\phi),\\
        f_7 &= 5\sqrt{\frac{3}{2}}\sin(2\theta)\sin(\phi),\quad
        f_8 = \frac{5(3 \cos (2 \theta )+1)}{2\sqrt{2}}.\\
    \end{split}
    \label{eq:angular-functions}
\end{equation}
These are the spherical harmonics $Y_l^m\left(\theta, \phi\right)$ with $l = 1, 2$, up to normalization and phase convention.
For the antiparticle decay $W^+ \to l^+ \nu_l$, CP invariance implies that the corresponding decay
density matrix takes the form
\begin{equation}
    \tilde{\rho}_{+} = \frac{1}{3}\left(\mathbb{1} + \sum_{i = 1}^{8} \bar{g}_i(\theta, \phi)R^i\right),
    \label{eq:Param-of-RhoP}
\end{equation}
where $\bar{g}_i(\theta, \phi) = g_i(\pi - \theta, \phi + \pi)$, and the associated angular
projectors are given by $\bar{f}_i(\theta, \phi) = f_i(\pi - \theta, \phi + \pi)$.

\subsection{Narrow Width Approximation}
\label{sec:DoublePoleApproximation}

Within the NWA, the full process in 
Eq.~(\ref{eq:full-process}) factorizes into production and decay subprocesses linked by 
the intermediate $W$ boson polarization states. The differential cross section 
$d\hat{\sigma}_{NWA}$ is given by
\begin{equation}
    d\hat{\sigma}_{NWA} = \frac{1}{2 s}\left(\frac{1}{\Gamma_W}\right)^2 
    \sum_{L \in \{LO, V, R, NP\}} \frac{1}{4} \operatorname{Tr}\left[\rho_{L} \cdot 
    \left(\tilde{\rho}_- \otimes \tilde{\rho}_+\right)\right] d\Pi_{L} \times 
    \left(3 d\Gamma_-\right) \left(3 d\Gamma_+\right),
    \label{eq:dsigmaD_compact}
\end{equation}
where $\Gamma_W$ denotes the total decay width of the $W$ boson. The differential partial 
decay width for $W^-\ (W^+)$ decay into a specific lepton flavor is given by
\begin{equation}
    d\Gamma_{\pm} = \frac{1}{2 m_W} \int \frac{1}{3} |\mathcal{M}_{\text{decay}}|^2 
    d\Pi_{\pm},
\end{equation}
where $d\Pi_{\pm}$ represents the 2-body phase space element of the respective decay. 
The factors of $\frac{1}{3}$ account for the average over $W$ boson polarization states.

Substituting the expansions of the production density matrices $\rho_L$ in 
Eq.~(\ref{eq:Param-of-RhoL}) and the normalized decay density matrices $\tilde{\rho}_-$ 
in Eq.~(\ref{eq:Param-of-RhoM}) as well as $\tilde{\rho}_+$ in Eq.~(\ref{eq:Param-of-RhoP}) into 
Eq.~(\ref{eq:dsigmaD_compact}), the differential cross section expands to
\begin{equation}
    \begin{split}
        d\hat{\sigma}_{NWA} = \frac{1}{2 s} \left(\frac{1}{\Gamma_W}\right)^2 
        \sum_{L \in \{LO, V, R, NP\}} \frac{1}{4} 
        \left( A^L + \sum_{i=1}^8 B_i^L g_i + \sum_{i=1}^8 \bar{B}_i^L \bar{g}_i + 
        \sum_{i,j=1}^8 C_{i,j}^L g_i \bar{g}_j \right) 
        d\Pi_L \, d\Gamma_- \, d\Gamma_+.
    \end{split}
    \label{eq:dsigmaD_expanded}
\end{equation}

To connect these theoretical density matrix components to experimentally measurable quantities, 
we exploit the orthogonality relations satisfied by the dual angular functions 
$f_i(\theta, \phi)$ and $\bar{f}_i(\theta, \phi)$ in Eq.~(\ref{eq:orthogonality}). Through 
these relations, the integrated polarization and correlation coefficients can be extracted 
directly from the decay angular distributions via moment projections:
\begin{equation}
    \hat{B}_i = \frac{1}{\hat{\sigma}_{NWA}}\int f_i \, d\hat{\sigma}_{NWA}, \quad
    \hat{\bar{B}}_i = \frac{1}{\hat{\sigma}_{NWA}}\int \bar{f}_i \, d\hat{\sigma}_{NWA}, 
    \quad \text{and} \quad
    \hat{C}_{i,j} = \frac{1}{\hat{\sigma}_{NWA}}\int f_i \bar{f}_j \, d\hat{\sigma}_{NWA},
    \label{eq:PolCosAndFiFj}
\end{equation}
where $\hat{\sigma}_{NWA} = \int d\hat{\sigma}_{NWA}$ is the total cross section under NWA. These 
projection relations demonstrate that the polarization parameters $\hat{B}_i$, $\hat{\bar{B}}_i$, 
and correlation coefficients $\hat{C}_{i,j}$ can be reconstructed experimentally by 
measuring the angular distributions of the final state charged leptons.

At this stage, one ambiguity remains concerning the labeling of the two identical
photons in the initial state. Since the two photons are physically
indistinguishable, the assignment of the labels $p_1$ and $p_2$ is purely
conventional.
Since this labeling enters the
definition of the local transverse axes in
Eqs.~(\ref{eq:hatnr1})~and~(\ref{eq:hatnr2}), exchanging the two photon labels generally
changes the polarization and correlation coefficients. 
Correspondingly, although the angular functions $f_i$ and $\bar f_i$ retain the same mathematical forms given in Eq.~(\ref{eq:angular-functions}), their values for a given event change because the lepton angles are then evaluated with respect to the transformed local axes.
Consequently, although individual polarization and correlation coefficients are
basis dependent, the moment relations in Eq.~(\ref{eq:PolCosAndFiFj}) retain the
same form. Once a convention for labeling the initial photons and defining the
local polarization axes is specified and used consistently, the coefficients
extracted from the measured lepton angular distributions can be directly compared
with the corresponding theoretical predictions. Different conventions therefore
provide equivalent basis descriptions of the same physical angular distribution.

In addition to full three dimensional angular projections, specific combinations of
correlation coefficients directly govern the azimuthal distributions of the final state 
charged leptons. Specifically, the differential distribution of the azimuthal angle difference 
$\Delta\phi_{\mathbf{v}} = \phi_{l^-} - \phi_{l^+}$ defined relative to the local reference frame axis 
$\mathbf{v}$ takes the general form
\begin{equation}
    \frac{1}{\sigma}\frac{d\sigma}{d\Delta\phi_{\mathbf{v}}} = \frac{1}{2\pi} +
    A_{\mathbf{v}} \cos(\Delta\phi_{\mathbf{v}}) + \tilde{A}_{\mathbf{v}}\sin(\Delta\phi_{\mathbf{v}}) +
    B_{\mathbf{v}} \cos(2\Delta\phi_{\mathbf{v}}) + \tilde{B}_{\mathbf{v}}\sin(2\Delta\phi_{\mathbf{v}}).
    \label{eq:dsigma_dDphi}
\end{equation}
For the process under consideration, since CP symmetry is preserved both within the SM (up to NLO EW if CKM matrix is taken to be identity matrix) and by the effective AQGC 
Lagrangian in Eq.~(\ref{eq:effective-lagrangian}), the CP-odd coefficients vanish identically, 
$\tilde{A}_{\mathbf{v}} = \tilde{B}_{\mathbf{v}} = 0$, for any choice of reference axis $\mathbf{v}$. 
Consequently, only the CP-even cosine modulation coefficients $A_{\mathbf{v}}$ and $B_{\mathbf{v}}$ are 
non-vanishing. Equations of the form in Eq.~(\ref{eq:dsigma_dDphi}) apply equally to fixed $\gamma\gamma$
CM energies ($\hat{\sigma}$) and to hadronic UPCs ($\sigma$).

When the reference axis $\mathbf{v}$ is chosen as the local $z$ axis in the respective $W$ boson 
rest frames ($\mathbf{v} = \hat{\mathbf{z}}$), the modulation coefficients evaluate to
\begin{equation}
    A_{\hat{\mathbf{z}}} = -\frac{3\pi}{128}\left(C_{5,5} + C_{6,6}\right),\quad
    B_{\hat{\mathbf{z}}} = \frac{1}{24\pi}\left(C_{2,2} + C_{3,3}\right).
    \label{eq:AzBz}
\end{equation}
Similarly, defining the azimuthal angle relative to the local transverse axes $\hat{\mathbf{x}}$ 
and $\hat{\mathbf{y}}$ yields
\begin{equation}
    \begin{split}
        A_{\hat{\mathbf{x}}} &= -\frac{3\pi}{128}\left(C_{1,1} + C_{5,5}\right),\\
        B_{\hat{\mathbf{x}}} &= \frac{1}{24\pi}\left(C_{7,7} + \frac{1}{4}(C_{3,3} + 3 C_{8,8}) -
        \frac{\sqrt{3}}{4}(C_{3,8} + C_{8,3})\right),
    \end{split}
    \label{eq:AxBx}
\end{equation}
and
\begin{equation}
    \begin{split}
        A_{\hat{\mathbf{y}}} &= -\frac{3\pi}{128}\left(C_{1,1} + C_{6,6}\right),\\
        B_{\hat{\mathbf{y}}} &= \frac{1}{24\pi}\left(C_{4,4} + \frac{1}{4}(C_{3,3} + 3 C_{8,8}) +
        \frac{\sqrt{3}}{4}(C_{3,8} + C_{8,3})\right).
    \end{split}
    \label{eq:AyBy}
\end{equation}
As shown in Appendix~\ref{app:MCE}, $A_{\mathbf{v}}$ and $B_{\mathbf{v}}$ receive no contributions 
from AQGCs for all three choices of $\mathbf{v}$ above. In fact, it can be demonstrated that 
these CP-even cosine modulations receive no AQGC contributions for an arbitrary reference axis 
$\mathbf{v}$. Consequently, measurements of these azimuthal correlations provide exceptionally clean 
benchmarks for precision tests of the SM.

\section{Numerical results}
\label{sec:Numerics}

In this section, we present numerical predictions for the total cross section
along with the polarization and correlation observables. Calculations are
performed in the on-shell renormalization scheme~\cite{Denner:1991kt, Denner:2019vbn}, using the physical pole
masses provided by the Particle Data Group (PDG)~\cite{ParticleDataGroup:2026mpi}.
For fermion loop corrections, only contributions from the third-generation
leptons ($\tau, \nu_{\tau}$) and the second- and third-generation quarks ($s, c,
b, t$) are included. The relevant input parameters are set to
\begin{equation}
    \begin{split}
        & m_W = 80.3625\ \mathrm{GeV},\quad m_Z = 91.1879\ \mathrm{GeV},\quad
        m_H = 125.13\ \mathrm{GeV},\\
        & m_{\tau} = 1.77693\ \mathrm{GeV},\quad m_c = 1.65\ \mathrm{GeV},\quad
        m_t = 172.60\ \mathrm{GeV},\\
        & m_s = 92.9\ \mathrm{MeV},\quad m_b = 4.788\ \mathrm{GeV}.
    \end{split}
\end{equation}
The weak mixing angle is fixed via $\cos \theta_W = \frac{m_W}{m_Z}$, and the
fine-structure constant is taken as $\alpha = \frac{1}{137}$.

Feynman diagrams and amplitudes are generated using
\textsc{FeynArts}~\cite{Hahn:2000kx} and symbolically evaluated with
\textsc{FeynCalc}~\cite{Shtabovenko:2023idz}. The resulting scalar one-loop
integrals are numerically computed using \textsc{LoopTools}~\cite{Hahn:1998yk},
and multi-dimensional phase space integrations are performed using the
\textsc{Cuba} library~\cite{Hahn:2004fe}.

Infrared divergences are regularized using the phase space slicing method. We
introduce a slicing parameter $x_{\mathrm{min}} = 10^{-3}$ to define the
soft photon region $E_{\gamma} < x_{\mathrm{min}} m_W$ in real emission
processes, with $E_{\gamma}$ denoting the energy of the final state photon in $\gamma\gamma$ CM frame.
We have explicitly verified that the combined sum of virtual and
real corrections remains stable against variations of $x_{\mathrm{min}}$ in the
range $10^{-8} < x_{\mathrm{min}} < 10^{-3}$.

Tensor integrals are handled using standard tensor integral decomposition (TID) and reduction method~\cite{Passarino:1978jh, Denner:1991kt}.
The intermediate Gram determinant, given by $\frac{1}{16}\beta^2 s^3 (1 - y^2)$,
vanishes near the production threshold ($\beta \to 0$) and in the forward/backward
scattering limits ($y^2 \to 1$). Although near these
phase space boundaries a dedicated expansion scheme is required~\cite{Denner:2005nn}, we regulate these
singular regions by applying a practical kinematic cut,
$\beta^2(1 - y^2) > \mathrm{TID}_{\mathrm{cut}}$, with a default value of
$\mathrm{TID}_{\mathrm{cut}} = 10^{-3}$. To ensure numerical stability, we verified that
varying $\mathrm{TID}_{\mathrm{cut}}$ down to $10^{-5}$ induces shifts below $0.1\%$
across almost all polarization and correlation observables presented later in
TABLE~\ref{table:Correlations1} except $C_{2,2}$, $C_{4,4}$ and $C_{6,6}$, with $C_{2,2}$ exhibiting a maximum variation of $0.24\%$.

\subsection{Results for fixed $\gamma\gamma$ CM energy}
\begin{figure}[h!]
    \centering
    \includegraphics[width=0.5\linewidth]{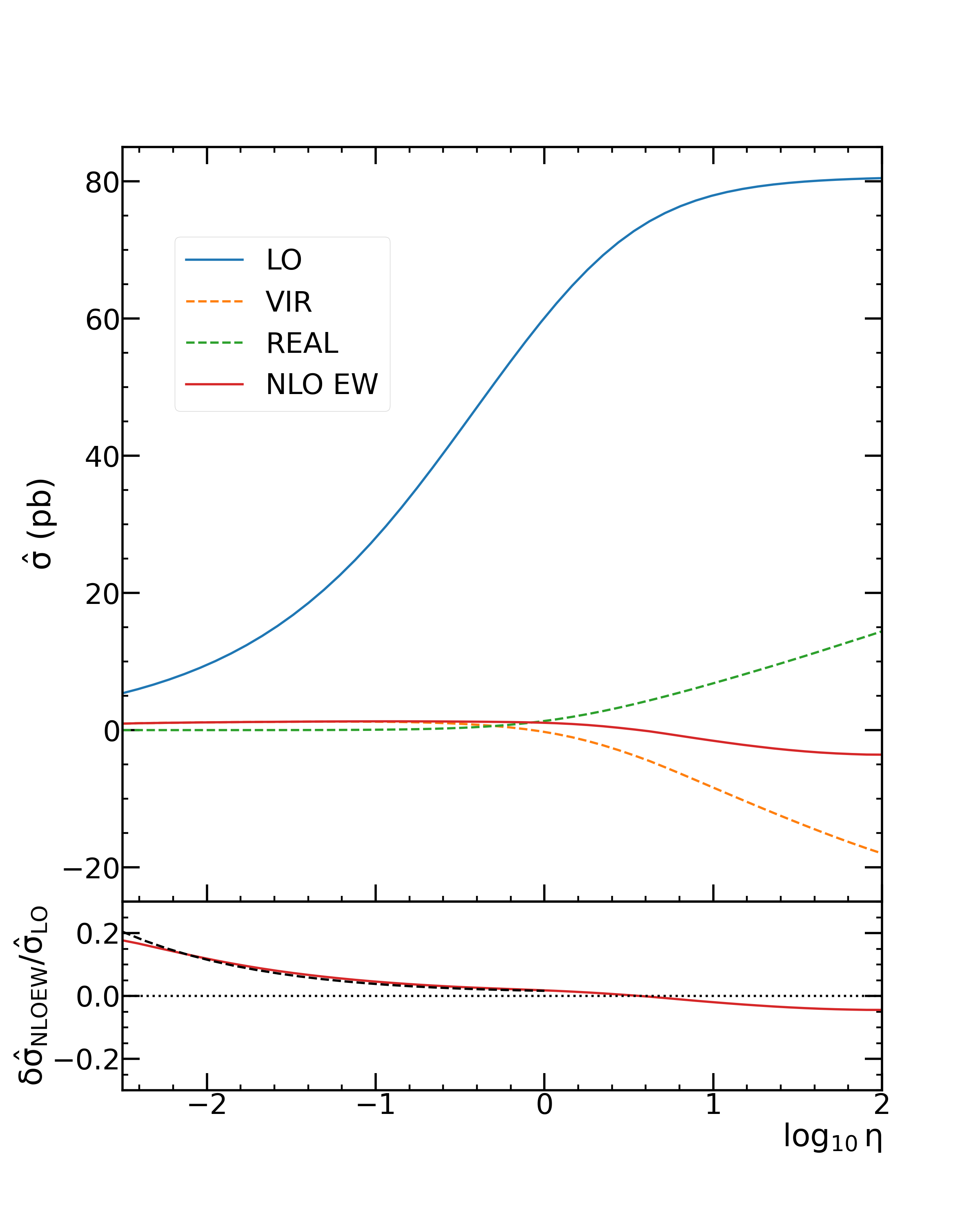}
    \caption{Total cross section $\hat{\sigma}$ for the 
    $\gamma\gamma\to W^-W^+$ process as a function of the threshold parameter
    $\eta = \frac{s}{4 m_W^2} - 1$ (upper panel). The solid blue line depicts
    the LO contribution, while the dashed orange and green lines represent the
    virtual and real NLO EW corrections, respectively. Their sum, labeled
    ``NLO EW'', is shown by the solid red line. The lower panel shows the
    relative NLO EW correction normalized to the LO cross section (solid red
    line). For comparison, the black dashed line displays the universal
    threshold approximation given in Eq.~(\ref{eq:threshold}), and the horizontal
    dotted black line at zero serves as a reference.}
    \label{fig:hard-cross-section}
\end{figure}

We first examine the case of fixed $\gamma\gamma$ CM energy.
FIG.~\ref{fig:hard-cross-section} displays the total cross section $\hat{\sigma}$
for the production process in Eq.~(\ref{eq:parton-cross-section}) as a function of $\log_{10}\eta$, where
$\eta = \frac{s}{4 m_W^2} - 1$ is a dimensionless kinematic threshold parameter. The
LO cross section grows monotonically with $\eta$, approaching
its asymptotic limit
$\lim\limits_{\eta\to\infty}\hat{\sigma}_{\mathrm{LO}} = \frac{8\pi\alpha^2}{m_W^2}
\approx 80.74\ \mathrm{pb}$ near $\log_{10}\eta = 2$. The absolute values of both virtual and real
corrections remain small and slowly varying for $\log_{10}\eta < 0$, but
increase rapidly in the region $\log_{10}\eta > 0$. However, their sum
displays a mild dependence on $\eta$, forming a stable plateau for
$\log_{10}\eta > 1.8$.

The relative NLO EW correction to the cross section becomes significant both
near the production threshold, where it reaches $+17.68\%$, and in the
high-$\eta$ regime, where it drops to $-4.45\%$. The behavior near threshold
is driven by the Coulomb singularity~\cite{Denner:1996wm}. While the LO cross section
vanishes near threshold, the NLO EW correction approaches a non-zero finite
value. Numerically we find $0.95\ \mathrm{pb}$ at $\log_{10}\eta = -2.5$, to be compared with the analytic limit $1.10\ \mathrm{pb}$ that follows from Eq.~(\ref{eq:threshold}) below. The difference is due to the kinematic cut $\beta^2(1-y^2)>\mathrm{TID}_{\mathrm{cut}}$, which at $\beta\approx 0.065$ excludes the region $|y|\gtrsim0.87$, about $13\%$ of the angular phase space from the one-loop contribution. The threshold behavior of the relative
NLO EW correction is universal and given by~\cite{Denner:1996wm}
\begin{equation}
    \frac{\delta\hat{\sigma}_{\mathrm{NLOEW}}}{\hat{\sigma}_{\mathrm{LO}}}
    = \frac{\hat{\sigma}_{\mathrm{NLOEW}} - \hat{\sigma}_{\mathrm{LO}}}
    {\hat{\sigma}_{\mathrm{LO}}}
    \approx \frac{\alpha\pi}{2\beta}
    = \frac{\alpha\pi}{2} \sqrt{1 + \frac{1}{\eta}}.
    \label{eq:threshold}
\end{equation}
As shown in the lower panel of FIG.~\ref{fig:hard-cross-section}, our full
calculation closely reproduces this universal threshold behavior. Minor
residual discrepancies near threshold arise from the approximate TID cutoff
scheme ($\mathrm{TID}_{\mathrm{cut}}$), which truncates the phase space near the
vanishing Gram determinant region ($\beta \to 0, y^2 \to 1$). Nevertheless,
the moderate size (a few percent) of the relative NLO EW corrections at high
energies indicates good perturbative stability over the considered energy range.

\begin{figure}[h!]
    \centering
    \includegraphics[width=0.48\textwidth]{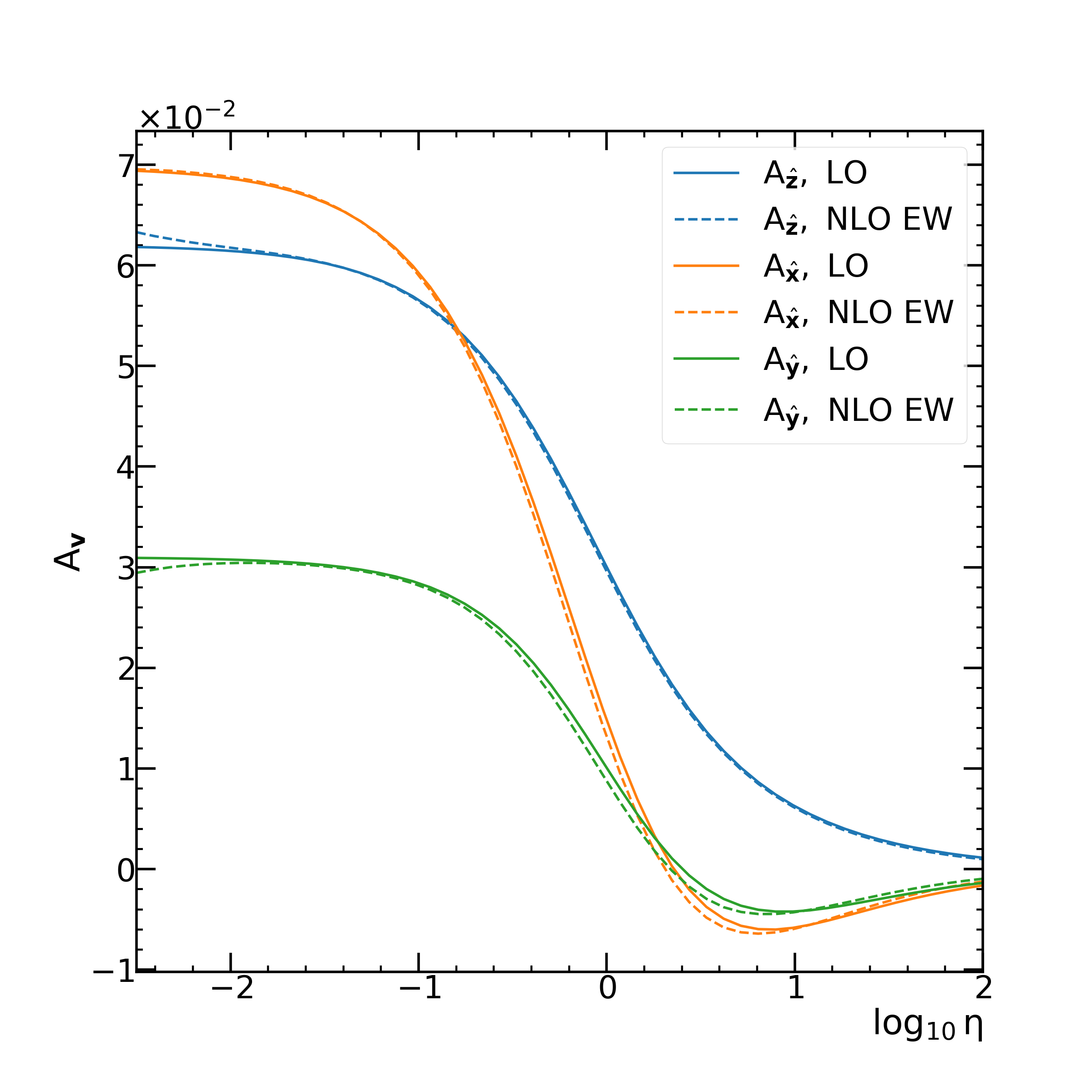}
    \includegraphics[width=0.48\textwidth]{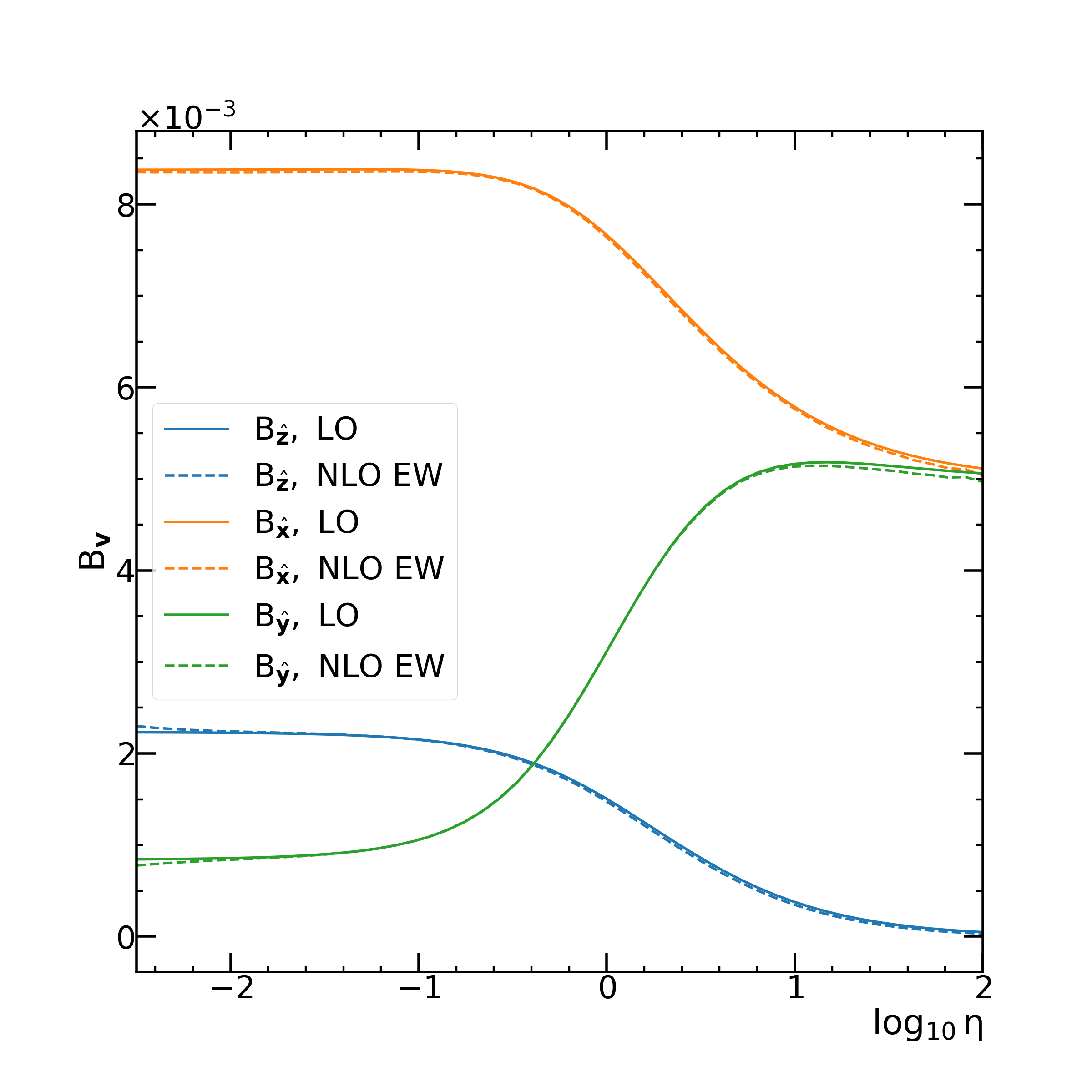}
    \caption{The azimuthal modulation coefficients $A_{\mathbf{v}}$ for
    $\cos(\Delta\phi_{\mathbf{v}})$ (left) and $B_{\mathbf{v}}$ for $\cos(2\Delta\phi_{\mathbf{v}})$ (right) as
    functions of $\log_{10}\eta$ for the 
    $\gamma\gamma\to W^-W^+$ process, evaluated for three choices of local
    rest frame reference axes: $\mathbf{v} \in \{\hat{\mathbf{z}}, \hat{\mathbf{x}}, \hat{\mathbf{y}}\}$. Solid
    lines represent the LO predictions, while dashed lines
    include NLO EW corrections.}
    \label{fig:AvBv}
\end{figure}

We now examine the azimuthal modulation coefficients across a broad range
of CM energies $\sqrt{s}$. FIG.~\ref{fig:AvBv} displays the energy dependence
of the $\cos(\Delta\phi_{\mathbf{v}})$ modulation amplitude $A_{\mathbf{v}}$
(left panel) and the $\cos(2\Delta\phi_{\mathbf{v}})$ modulation amplitude
$B_{\mathbf{v}}$ (right panel) for the three local reference frame axes
$\mathbf{v} \in \left\{\hat{\mathbf{z}}, \hat{\mathbf{x}}, \hat{\mathbf{y}}\right\}$. As derived in
Eqs.~(\ref{eq:AzBz})--(\ref{eq:AyBy}), these modulation parameters are directly determined by
specific linear combinations of the production correlation coefficients
$C_{i,j}$.

Across the entire energy spectrum, the NLO EW corrections to both
$A_{\mathbf{v}}$ and $B_{\mathbf{v}}$ induce only mild quantitative shifts
relative to their LO baselines, demonstrating excellent perturbative stability
for these azimuthal observables. The single-angle amplitudes
$A_{\mathbf{v}}$ start from their maximum values near threshold
($\log_{10}\eta \to -\infty$) and fall off toward zero in the
high energy limit ($\log_{10}\eta \to \infty$). In contrast, the double-angle
coefficients $B_{\mathbf{v}}$ exhibit distinct qualitative behaviors depending
on the spatial orientation of the reference axis: while $B_{\hat{\mathbf{z}}}$ and
$B_{\hat{\mathbf{x}}}$ decrease monotonically toward their high-energy limits,
$B_{\hat{\mathbf{y}}}$ rises monotonically above threshold, with both transverse
components ($B_{\hat{\mathbf{x}}}$ and $B_{\hat{\mathbf{y}}}$) approaching the shared asymptotic
LO value of $\frac{1}{64\pi}$ as $\log_{10}\eta \to \infty$.

In addition to observables already present at LO in SM, another 
crucial class of correlation coefficients vanishes identically at LO 
in the SM while remaining directly sensitive to AQGCs, as detailed 
in Appendix~\ref{app:MCE}. To establish a reliable baseline for 
future new physics searches, it is essential to determine their 
SM predictions beyond LO.
We restrict our numerical analysis to $C_{1,2}$, $C_{4,5}$, and 
$C_{6,7}$, as they are the only AQGC sensitive correlation 
observables that survive phase space integration to yield non-zero 
integrated values. Specifically, $C_{1,2}$ probes $a_c$ 
exclusively, whereas $C_{4,5}$ and $C_{6,7}$ depend on both $a_0$ 
and $a_c$. 

\begin{figure}[h!]
    \centering
    \includegraphics[width=0.5\linewidth]{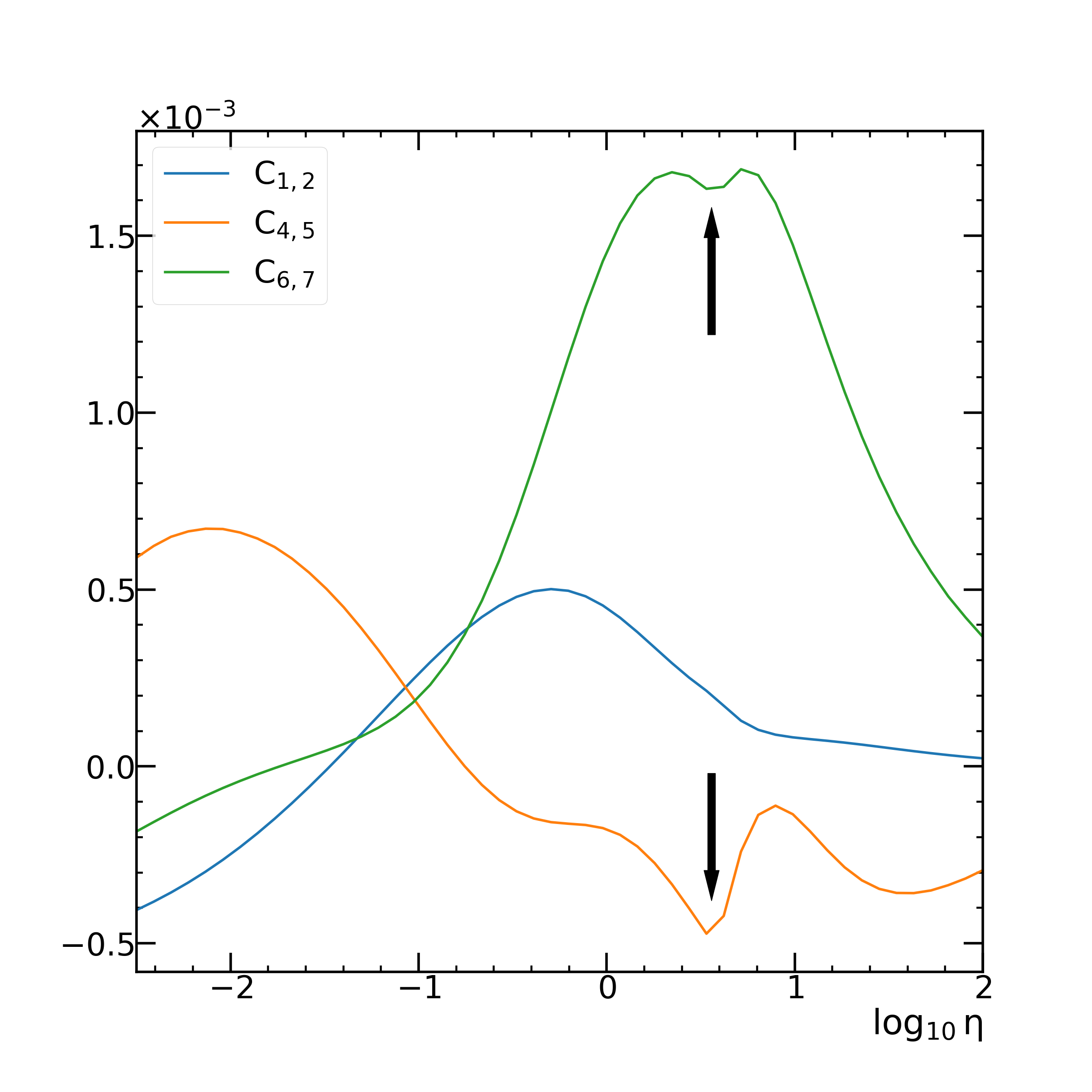}
    \caption{Correlation coefficients $C_{1,2}$, $C_{4,5}$, and $C_{6,7}$ as
    functions of $\log_{10}\eta$ at NLO EW accuracy. These coefficients vanish identically at LO. The black arrows indicate
    turning points at the $t\bar{t}$ threshold ($\sqrt{s} = 2m_t$).}
    \label{fig:C_AQGC}
\end{figure}

As shown in FIG.~\ref{fig:C_AQGC}, NLO EW corrections 
break this LO zero-baseline property by generating non-zero SM 
values across the energy spectrum, with $C_{6,7}$ reaching a peak 
magnitude near $1.65\times 10^{-3}$. The turning points near 
$\log_{10}\eta \approx 0.558$ ($\sqrt{s} = 2m_t$, indicated by black 
arrows) mark the threshold onset of top quark pair production in 
virtual loops. Accounting for these NLO EW baseline shifts is 
essential in high precision experimental searches to avoid 
misinterpreting SM loop effects as genuine AQGC signals.

\subsection{Results in proton-proton UPCs}

\begin{figure}[h!]
    \includegraphics[width=0.5\linewidth]{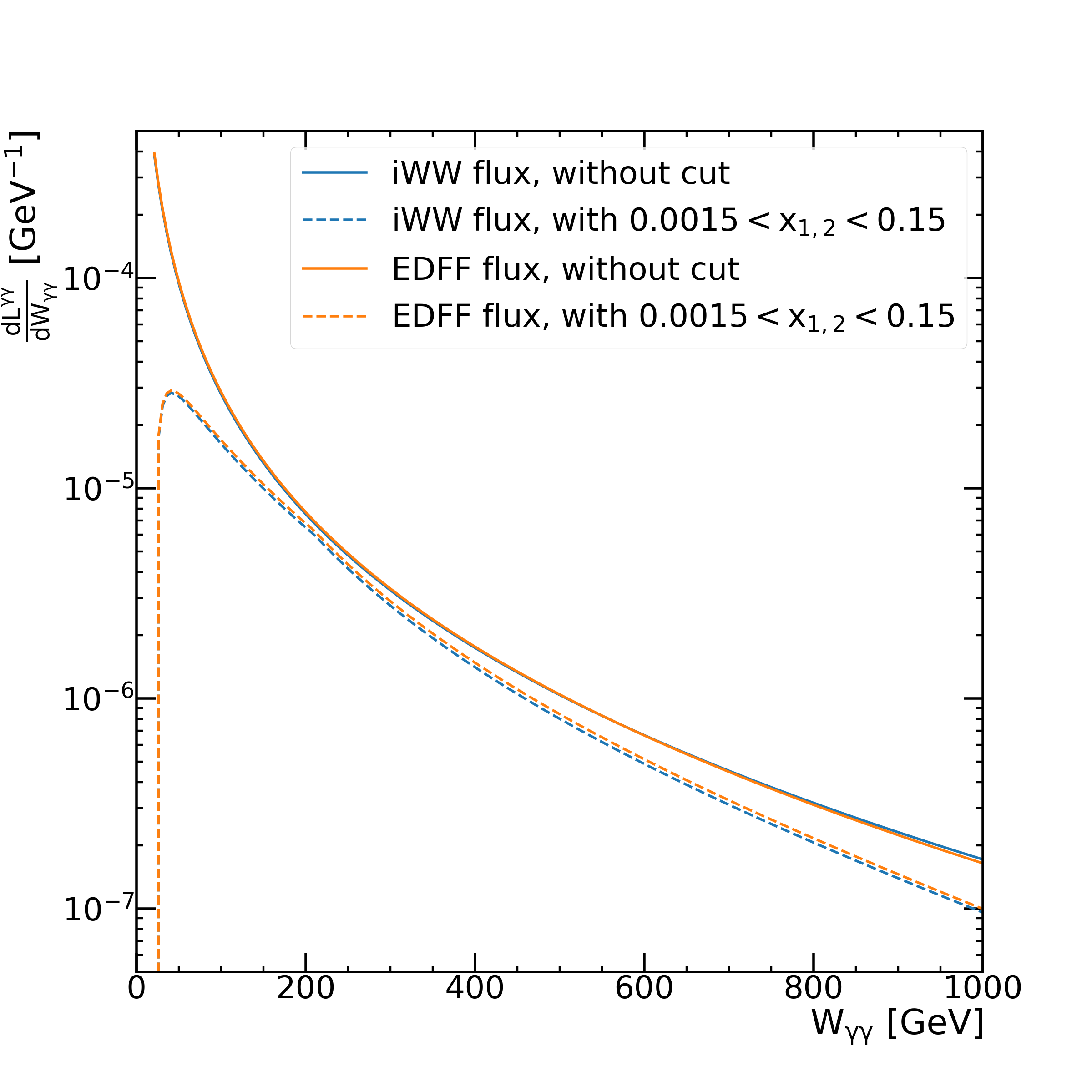}
    \caption{Two-photon luminosity spectrum $\frac{dL^{\gamma\gamma}}{dW_{\gamma\gamma}}$ as
    a function of the $\gamma\gamma$ CM energy $W_{\gamma\gamma}$. The
    variables $x_{1,2}$ represent the beam momentum fractions carried by the
    initial state photons. Predictions using the improved Weizs\"acker-Williams
    (iWW) approximation and the electric dipole form factor (EDFF) are shown in
    blue and orange, respectively. Solid lines denote results without kinematic
    cuts, whereas dashed lines incorporate the forward proton acceptance cuts
    $0.0015 < x_{1,2} < 0.15$.}
    \label{fig:pdf}
\end{figure}

We now extend our analysis to realistic $pp$ UPCs by convolving the $\gamma\gamma$ cross section with the
effective photon flux. FIG.~\ref{fig:pdf} displays the two-photon luminosity,
$\frac{dL^{\gamma\gamma}}{dW_{\gamma\gamma}}$, as a function of the $\gamma\gamma$
CM energy $W_{\gamma\gamma} = \sqrt{s}$. Predictions are shown for
two flux models: the improved Weizs\"acker-Williams (iWW)
approximation~\cite{Frixione:1993yw} and the electric dipole form factor
(EDFF) parameterization~\cite{Baltz:2007kq}. Detailed formulations for these
photon spectra can be found in the corresponding references and are omitted
here for brevity. Here, $x_{1,2}$ denote the beam energy fractions carried by
the incoming photons with momenta $p_{1,2}$.
The forward-proton acceptance cut $0.0015 < x_{1,2} < 0.15$ reflects typical detector acceptance windows~\cite{Chapon:2009hh}. Forward proton tagging is vital for
kinematically reconstructing the $W^-W^+$ CM frame in the presence
of invisible final state neutrinos. Our photon spectrum implementation has been
validated against earlier benchmarks in
Refs.~\cite{Pierzchala:2008xc,Chapon:2009hh}.

By convolving the partonic cross section with the two photon luminosity, we can
evaluate both the total cross section
\begin{equation}
    \sigma =
    \int d W_{\gamma\gamma} \frac{d L^{\gamma\gamma}}{d W_{\gamma\gamma}} \hat{\sigma}(s=W^2_{\gamma\gamma})
\end{equation}
for $pp \to p W^-W^+ p$, with $\hat{\sigma}$ given in Eq.~(\ref{eq:parton-cross-section}),
and the associated polarization and correlation observables in $pp$
UPCs.
Since numerical differences between the iWW and EDFF photon fluxes are
negligible, all subsequent numerical results are presented using the iWW flux.

TABLE~\ref{table:ppCrossSectionNoCut} summarizes the integrated cross sections
for $\gamma\gamma$ induced $W^- W^+$ production in $pp$ UPCs across four
representative LHC CM energies ($\sqrt{s_{pp}} = 8$, $13$, $13.6$,
and $14\text{ TeV}$). Note that AQGCs do
not alter the unpolarized cross section. As expected, the integrated
cross section increases significantly with collision energy, more than doubling
from $44.23\text{ fb}$ at $8\text{ TeV}$ to $95.69\text{ fb}$ at $14\text{ TeV}$
at LO. Across all evaluated energies, the net relative NLO EW corrections ($\delta$)
remain positive but small, staying below $0.5\%$ and monotonically decreasing from
$+0.47\%$ at $8\text{ TeV}$ to $+0.04\%$ at $14\text{ TeV}$.
\begin{table}[h!]
\begin{ruledtabular}
    \begin{tabular}{cccc}
        $\sqrt{s_{pp}}\ (\mathrm{TeV})$&$\sigma_{\mathrm{LO}}\ (\mathrm{fb})$&$\sigma_{\mathrm{NLO EW}}\ (\mathrm{fb})$&$\delta=\left(\sigma_{\mathrm{NLO EW}} - \sigma_{\mathrm{LO}}\right)/\sigma_{\mathrm{LO}}$\\
        \hline
        8 & 44.23 & 44.44 & 0.47\%\\
        13 & 87.03 & 87.12 & 0.10\%\\
        13.6 & 92.23 & 92.29 & 0.07\%\\
        14 & 95.69 & 95.73 & 0.04\%\\
    \end{tabular}
    \caption{Total cross sections for $\gamma\gamma$ induced $W^- W^+$ production in
    $pp$ UPCs across various CM
    energies $\sqrt{s_{pp}}$. Listed are the LO results $\sigma_{\mathrm{LO}}$
    (second column), predictions at NLO EW accuracy $\sigma_{\mathrm{NLOEW}}$
    (third column), and the corresponding relative NLO EW corrections $\delta$
    (fourth column). Results are computed using the photon flux in the iWW
    approximation. No forward proton acceptance cuts are applied in obtaining these
    values.}
    \label{table:ppCrossSectionNoCut}
\end{ruledtabular}
\end{table}

Our LO cross section prediction at $14\text{ TeV}$ shows good agreement with the
result reported in Ref.~\cite{Chapon:2009hh}. In Ref.~\cite{Shao:2025bma}, NLO EW
corrections were evaluated using a charge form factor photon flux at identical LHC
energies. Compared to their results, our LO cross section is systematically larger
by a factor of $1.37$--$1.45$, while our relative NLO EW corrections correspond to
$40\%$--$68\%$ of their reported values. These discrepancies originate from the proton survival probability, the
different photon flux parameterizations employed as well as the simplified
treatment of tensor integral reduction near phase space boundaries in our current
setup. Crucially, the flux induced normalization uncertainties are expected to
largely cancel in the ratios defining the polarization and correlation observables
in Eq.~\eqref{eq:PolCosAndFiFj}.

\begin{figure}[h!]
    \centering
    \includegraphics[width=0.5\linewidth]{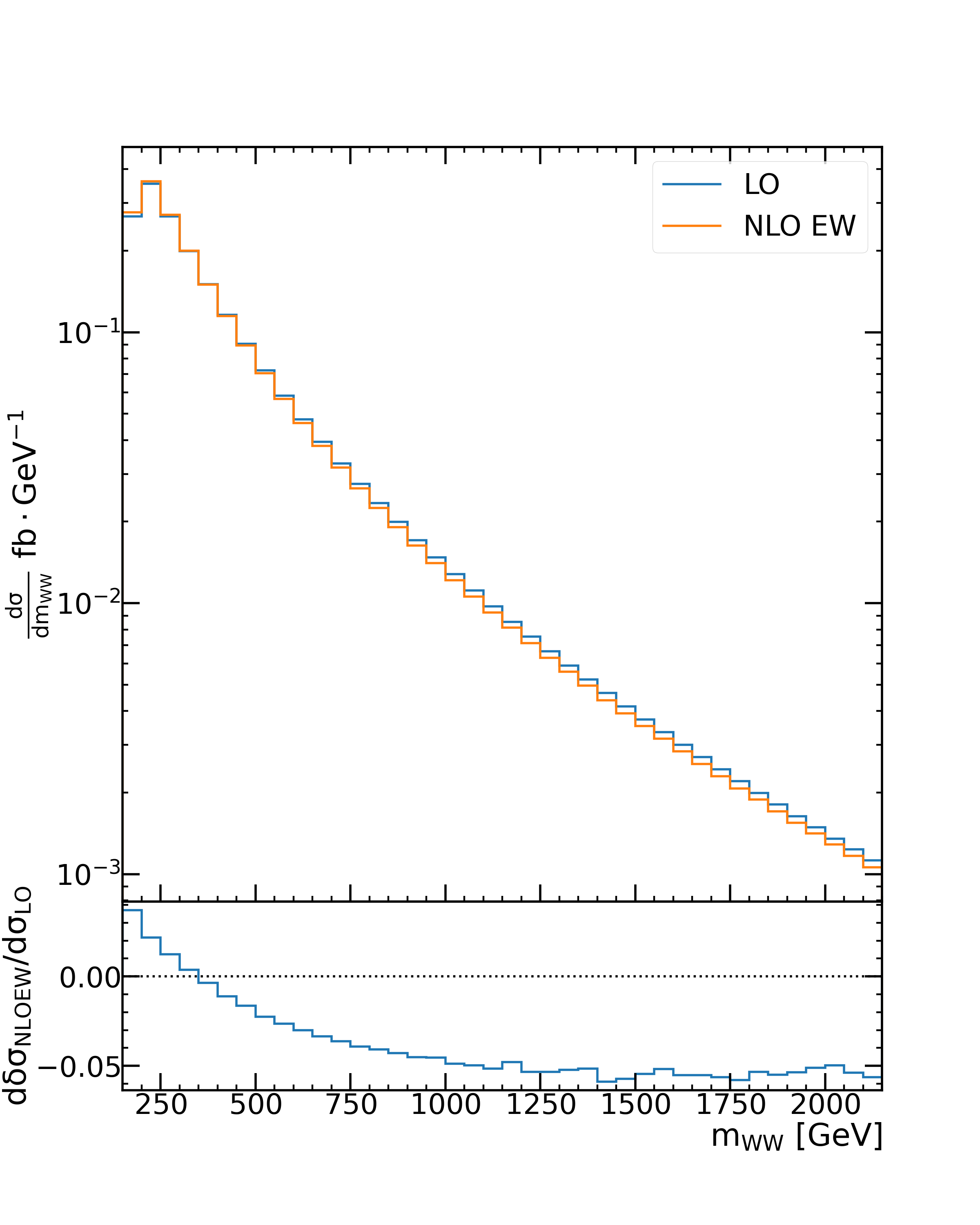}
    \caption{Differential distribution with respect to the $W$ boson pair
    invariant mass $m_{WW}$ in $pp$ UPCs at
    $\sqrt{s_{pp}} = 14\ \mathrm{TeV}$, computed using the iWW photon flux
    without forward proton acceptance cuts. The upper panel shows the LO
    prediction (blue) and the result with NLO EW corrections included
    (orange). The lower panel displays the relative NLO EW correction
    $\mathrm{d}\delta\sigma_{\mathrm{NLOEW}}/\mathrm{d}\sigma_{\mathrm{LO}}$
    as a solid blue line, while a horizontal
    dotted black line at zero serves as a reference.}
    \label{fig:dsigma_dmww}
\end{figure}

Besides inclusive total cross sections, we also investigate the
differential distribution with respect to the invariant mass $m_{WW}$ of
the $W$ boson pair in $pp$ UPCs at $\sqrt{s_{pp}} = 14\ \mathrm{TeV}$,
presented in FIG.~\ref{fig:dsigma_dmww}. As shown in the upper panel, the cross section
decreases rapidly across two orders of magnitude as $m_{WW}$ increases
from threshold up to $2.15\ \mathrm{TeV}$, with the LO baseline
(solid blue) and the result including NLO EW corrections (solid orange)
remaining visually close throughout the spectrum. The lower panel
highlights the relative NLO EW correction $\mathrm{d}\delta\sigma_{\mathrm{NLOEW}}
/ \mathrm{d}\sigma_{\mathrm{LO}}$. Near the production threshold ($m_{WW}
\approx 2 m_W$), the relative correction is positive ($\approx +4\%$)
owing to the Coulomb enhancement. As $m_{WW}$ grows, the relative NLO EW
correction steadily decreases, crossing zero and turning negative near $m_{WW} \sim 350\
\mathrm{GeV}$, eventually stabilizing around
$-5\%$ to $-7\%$ in the high invariant mass region ($m_{WW} \sim 2.15\
\mathrm{TeV}$).

We now turn to the polarization and correlation observables. Out of the 16 polarization
($B_i, \bar{B}_i$) and 64 correlation ($C_{i,j}$) coefficients, only a subset of them yields
non-zero integrated values at LO in the SM. As a consequence of CP symmetry, the relations
$\bar{B}_i = B_i$ and $C_{j,i} = C_{i,j}$ hold, allowing us to report only the independent
components $B_i$ and $C_{i,j}$ with $i \le j$.
TABLE~\ref{table:Correlations1} compiles the integrated predictions for those
independent coefficients that possess a non-zero net value at LO, alongside the
corresponding NLO EW predictions and the relative shifts $\delta$, evaluated for
$pp$ UPCs at $\sqrt{s_{pp}} = 14~\mathrm{TeV}$. Note that
forward proton tagging is required for full kinematic reconstruction of the event;
accordingly, a forward proton acceptance cut of $0.0015 < x_{1,2} < 0.15$ is applied,
motivated by Refs.~\cite{Pierzchala:2008xc,Chapon:2009hh}.

Sizable single boson polarizations are induced, dominated by $B_8$ which reaches $0.533$
at LO, with NLO EW corrections to $B_i$ remaining well under $1\%$. Since $R^8=(3S_z^2-2\,\mathbb{1})/\sqrt{2}$, the coefficient $B_8$ measures the longitudinal polarization fraction $f_0$ of the $W^-$ boson through $B_8=(1-3f_0)/\sqrt{2}$, ranging from $B_8=0$ for an unpolarized $W$ to $B_8=1/\sqrt{2}$ for a purely transverse one. The value $B_8\approx0.533$ corresponds to $f_0\approx8\%$, reflecting the dominance of transversely polarized $W$ bosons. Among the correlation
parameters, $C_{8,8}$ exhibits the largest absolute magnitude ($0.428$ at LO), while
$C_{2,2}$, the smallest coefficient non-vanishing at LO, receives the largest relative
NLO EW shift of $-33.84\%$, followed by $C_{1,1}$ at $+14.37\%$. For the remaining
correlation observables present at LO, NLO EW corrections range from sub-percent to a few
percent, establishing a reliable baseline for SM precision tests.
\begin{table}[h!]
\begin{ruledtabular}
\begin{tabular}{cccc}
    $\mathcal{O}$ & $\mathcal{O}_{\mathrm{LO}}$ & $\mathcal{O}_{\mathrm{NLO EW}}$ & $\delta$\\
    \hline
    $B_3$ & $8.43\times 10^{-2}$ & $8.45\times 10^{-2}$ & $0.25\%$\\
    $B_8$ & $5.33\times 10^{-1}$ & $5.31\times 10^{-1}$ & $-0.38\%$\\
    $C_{1,1}$ & $5.54\times 10^{-2}$ & $6.34\times 10^{-2}$ & $14.37\%$\\
    $C_{2,2}$ & $1.26\times 10^{-3}$ & $8.34\times 10^{-4}$ & $-33.84\%$\\
    $C_{3,3}$ & $7.76\times 10^{-2}$ & $7.68\times 10^{-2}$ & $-1.00\%$\\
    $C_{3,8}$ & $-6.50\times 10^{-3}$ & $-6.77\times 10^{-3}$ & $4.20\%$\\
    $C_{4,4}$ & $-3.87\times 10^{-2}$ & $-4.01\times 10^{-2}$ & $3.52\%$\\
    $C_{5,5}$ & $-1.68\times 10^{-1}$ & $-1.68\times 10^{-1}$ & $-0.06\%$\\
    $C_{6,6}$ & $-1.19\times 10^{-1}$ & $-1.19\times 10^{-1}$ & $0.27\%$\\
    $C_{7,7}$ & $1.73\times 10^{-1}$ & $1.73\times 10^{-1}$ & $0.24\%$\\
    $C_{8,8}$ & $4.28\times 10^{-1}$ & $4.26\times 10^{-1}$ & $-0.28\%$\\
\end{tabular}
\end{ruledtabular}
    \caption{Predictions for polarization and correlation coefficients with non-zero net values
    at LO, along with their corresponding NLO EW corrections, for $pp$ collisions at
    $\sqrt{s_{pp}} = 14~\mathrm{TeV}$. The relative NLO EW corrections are denoted by
    $\delta = (\mathcal{O}_{\mathrm{NLOEW}} - \mathcal{O}_{\mathrm{LO}})/\mathcal{O}_{\mathrm{LO}}$.
    Results are computed using the photon flux in the iWW
    approximation, subject to forward proton acceptance cut of $0.0015 < x_{1,2} < 0.15$.}
    \label{table:Correlations1}
\end{table}

\begin{figure}[h!]
    \centering
    \includegraphics[width=0.5\linewidth]{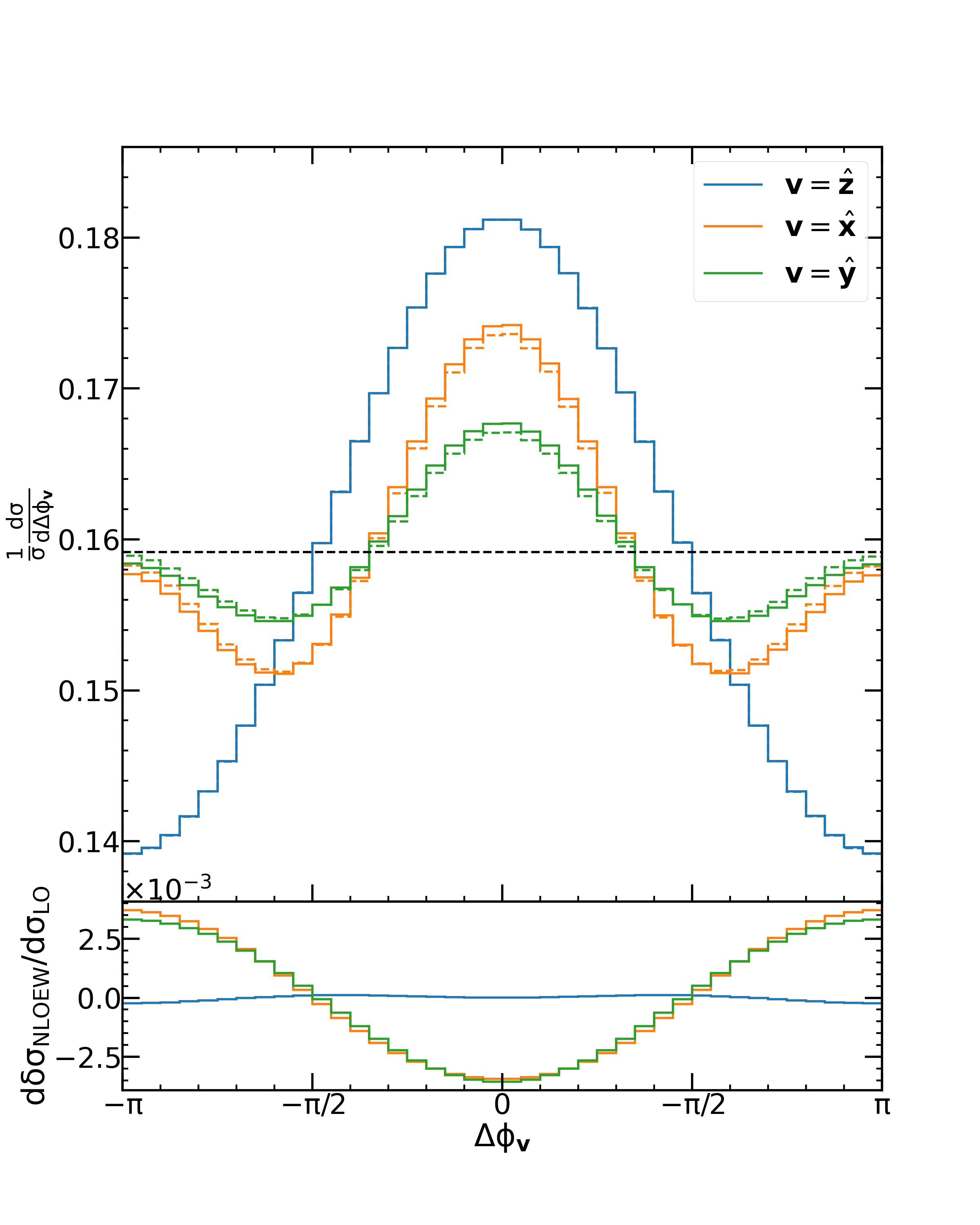}
    \caption{Azimuthal correlation distributions 
    $\frac{1}{\sigma}\frac{d\sigma}{d\Delta\phi_{\mathbf{v}}}$ 
    for three choices of local reference frame axes 
    ($\mathbf{v} \in \{\hat{\mathbf{z}}, \hat{\mathbf{x}}, \hat{\mathbf{y}}\}$). 
    Upper panel: LO predictions (solid lines) compared with predictions 
    including NLO EW corrections (dashed lines). Lower panel: Relative NLO EW corrections 
    with respect to LO predictions, defined as 
    $(d\sigma_{\mathrm{NLOEW}} - d\sigma_{\mathrm{LO}})/d\sigma_{\mathrm{LO}}$. 
    All results are evaluated for $pp$ UPCs at $\sqrt{s_{pp}} = 14\mathrm{~TeV}$ 
    using the iWW photon flux with forward proton acceptance cut 
    $0.0015 < x_{1,2} < 0.15$, consistent with the setup in TABLE~\ref{table:Correlations1}.}
    \label{fig:dsigma_dDphi}
\end{figure}

Having established the polarization and correlation coefficients in TABLE~\ref{table:Correlations1}, 
we now turn our attention to the differential structure of the final state leptons. 
FIG.~\ref{fig:dsigma_dDphi} illustrates the differential distributions (upper panel) 
and relative NLO EW corrections (lower panel) for the normalized azimuthal angle difference 
$\frac{1}{\sigma}\frac{d\sigma}{d\Delta\phi_{\mathbf{v}}}$ of the final state charged 
leptons in $pp$ UPCs at $\sqrt{s_{pp}} = 14\text{ TeV}$. The distributions 
are evaluated using the iWW photon flux with forward proton tagging cut 
$0.0015 < x_{1,2} < 0.15$ across three distinct local rest frame reference axes 
$\mathbf{v} \in \{\hat{\mathbf{z}}, \hat{\mathbf{x}}, \hat{\mathbf{y}}\}$. 
The upper panel demonstrates that the azimuthal modulations depend strongly on the selected 
orientation axis $\mathbf{v}$, with $\cos(2\Delta\phi_{\mathbf{v}})$ being most visible 
for the transverse axes $\mathbf{v}=\hat{\mathbf{x}}$ and $\mathbf{v}=\hat{\mathbf{y}}$.

The lower panel shows that NLO EW radiative corrections induce only a mild quantitative 
impact on the overall shape of these azimuthal correlations. 
For $\mathbf{v}=\hat{\mathbf{z}}$, the relative correction is smallest across the 
entire range of $\Delta\phi_{\hat{\mathbf{z}}}$. For the transverse axes 
($\mathbf{v}=\hat{\mathbf{x}}$ and $\mathbf{v}=\hat{\mathbf{y}}$), the relative NLO EW corrections 
exhibit small, symmetric oscillations bounded within $\pm 0.40\%$, reaching their maximum 
enhancement around $\Delta\phi_{\mathbf{v}} = \pm\pi$ and minimum near 
$\Delta\phi_{\mathbf{v}} = 0$. Because these CP-even azimuthal modulations receive zero 
contribution from AQGCs in Eq.~(\ref{eq:effective-lagrangian}) for any reference axis 
$\mathbf{v}$, their extreme perturbative stability under NLO EW corrections establishes 
them as exceptionally clean, robust benchmarks for precision tests of the SM.

Finally, we turn to analyzing the correlation coefficients that exhibit direct sensitivity 
to AQGCs.
In TABLE~\ref{table:Correlations2}, we present predictions for correlation coefficients 
that vanish at LO but display sensitivity to AQGCs. Specifically, $C_{1,2}$ is sensitive 
exclusively to $a_c$, whereas $C_{4,5}$ and $C_{6,7}$ are sensitive to both $a_0$ and 
$a_c$. To induce a correlation of $C_{1,2} = 2.70 \times 10^{-4}$, our calculation 
indicates that $a_c$ must reach $|a_c|/\Lambda^2 > 6.74 \times 10^{-6}\text{ GeV}^{-2}$. 
Similarly, to generate $C_{4,5} = 2.28 \times 10^{-4}$, either 
$|a_0|/\Lambda^2 > 4.48 \times 10^{-8}\text{ GeV}^{-2}$ or 
$|a_c|/\Lambda^2 > 2.19 \times 10^{-7}\text{ GeV}^{-2}$ is required. The corresponding 
bounds deduced from $C_{6,7}$ are $|a_0|/\Lambda^2 > 2.62 \times 10^{-7}\text{ GeV}^{-2}$ 
and $|a_c|/\Lambda^2 > 9.01 \times 10^{-7}\text{ GeV}^{-2}$.

\begin{figure}[h!]
    \includegraphics[width=0.5\linewidth]{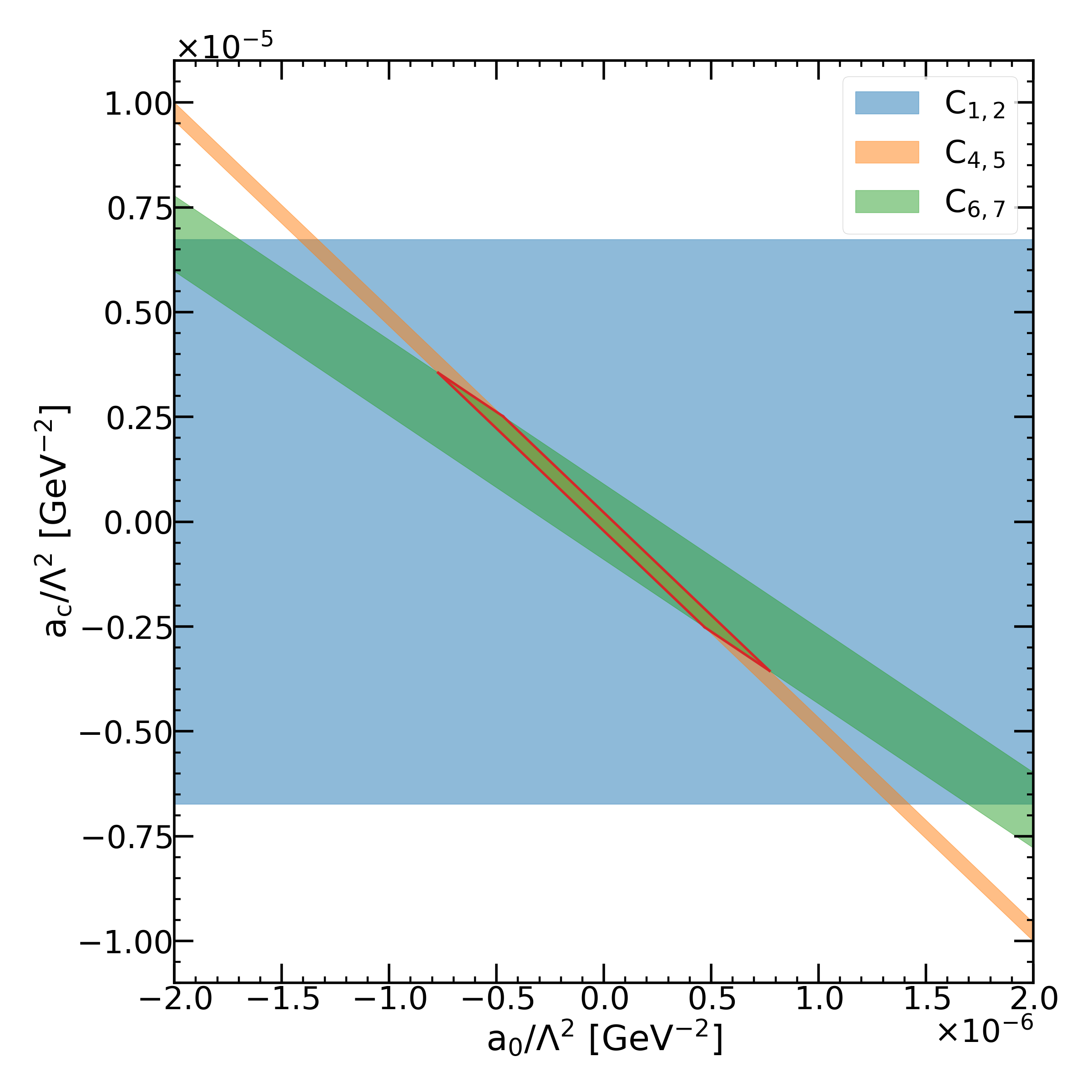}
    \caption{Sensitivity thresholds in the $(a_0/\Lambda^2,a_c/\Lambda^2)$
    plane derived by comparing the AQGC contributions with the SM NLO EW
    contributions to $C_{1,2}$, $C_{4,5}$, and $C_{6,7}$. The region enclosed by the red boundary represents 
    the overlapping region of all three constraint bands, defining the parameter regime where 
    NLO EW corrections must be incorporated into SM predictions to extract reliable AQGC 
    limits.}
    \label{fig:sensitivity}
\end{figure}

In FIG.~\ref{fig:sensitivity}, we project these values onto the two-dimensional
$(a_0/\Lambda^2,a_c/\Lambda^2)$ parameter space. Each band denotes the region
in which the magnitude of the AQGC contribution to the corresponding correlation
coefficient is smaller than its SM one-loop contribution. The region enclosed by
the red lines represents the overlap of the three bands, within which the SM
loop-induced contributions to all three coefficients exceed the corresponding
AQGC contributions. Once the experimental precision on these coefficients reaches
the $10^{-4}$--$10^{-3}$ level, the SM loop-induced baselines must therefore be
included for a reliable interpretation in terms of AQGCs.

For comparison, the current constraints on $a_0$ and $a_c$ reported in
Ref.~\cite{CMS:2022dmc} extend beyond the parameter range displayed in
FIG.~\ref{fig:sensitivity}. However, these constraints are derived from
high-mass $\gamma\gamma\to W^-W^+$ production rather than from the correlation
observables considered here. They should therefore be regarded only as an
indication of the coupling range currently probed experimentally, rather than
as a direct comparison with our sensitivity bands.

\begin{table}[h!]
\begin{ruledtabular}
\begin{tabular}{ccc}
    $\mathcal{O}$ & $\mathcal{O}_{\mathrm{LO}}$ & $\mathcal{O}_{\mathrm{NLOEW}}$\\
    \hline
    $C_{1,2}$ & 0 & $2.70\times 10^{-4}$\\
    $C_{4,5}$ & 0 & $2.28\times 10^{-4}$\\
    $C_{6,7}$ & 0 & $1.33\times 10^{-3}$\\
\end{tabular}
\end{ruledtabular}
    \caption{Predictions for correlation coefficients $C_{1,2}$, $C_{4,5}$, 
    and $C_{6,7}$ in $pp$ UPCs at $\sqrt{s_{pp}} = 14\text{ TeV}$. These 
    observables vanish at LO in the SM but are sensitive to 
    AQGCs. Results are calculated using iWW photon flux with forward proton acceptance cut of 
    $0.0015 < x_{1,2} < 0.15$.}
    \label{table:Correlations2}
\end{table}

\section{SUMMARY}
\label{sec:Summary}

In this paper, we studied NLO EW corrections to $\gamma\gamma$-induced
$W^-W^+$ production in $pp$ UPCs at the LHC within the NWA, with the
$W$-boson decays treated at LO. Using an orthogonal set of $3\times3$
traceless Hermitian matrices $R^i$ to parametrize the production and decay
density matrices, we constructed the corresponding dual angular functions
$f_i(\theta,\phi)$. These functions allow the polarization coefficients
$B_i$ and $\bar B_i$ and the correlation coefficients $C_{i,j}$ to be
extracted directly from the leptonic decay angular distributions through
moment projections.

At the hard-scattering level, NLO EW corrections enhance the cross section
near the $W^+W^-$ threshold through Coulomb effects, while reducing it by
up to $4.45\%$ at higher energies. After convolution with the photon flux
in $pp$ UPCs at $\sqrt{s_{pp}}=14$ TeV, the correction to the total
unpolarized cross section is small and positive, about $+0.04\%$. The NLO EW corrections to the
invariant-mass distribution exhibit a pronounced energy dependence, ranging from about $+4\%$ near threshold to
$-5\%$~--~$-7\%$ in the high-mass region around $m_{WW}\sim2.15$ TeV.
In contrast, the normalized azimuthal correlation distributions receive
corrections below $0.40\%$. Individual correlation coefficients can,
however, exhibit much larger relative corrections, reaching $-33.84\%$
for $C_{2,2}$.

We further investigated correlation observables that are sensitive to
AQGCs. In the SM, the coefficients $C_{1,2}$, $C_{4,5}$, and $C_{6,7}$
vanish at LO but acquire non-zero values through NLO EW corrections. At
$\sqrt{s_{pp}}=14$ TeV, these loop-induced contributions are of order
$10^{-4}$--$10^{-3}$. By comparing them with the corresponding AQGC
contributions, we identified the region of the
$(a_0/\Lambda^2,a_c/\Lambda^2)$ parameter space in which the SM
loop-induced baselines become relevant for the interpretation of these
correlation observables. Our results provide a quantitative assessment of
the role of NLO EW effects in future precision studies of
polarization and correlation effects of $W$ boson pair produced in $\gamma\gamma$ annihilation.

\section*{Acknowledgements}
The authors thank
members of the Institute of Theoretical Physics at Shandong University,
especially Haitao Li and Kirill Kudashkin, for their helpful communications.
This work is supported in part by National Natural Science Foundation of China under
the Grants No. 12235008, No. 12321005, No. 12475083, and No. 12405121.

\begin{appendix}
\section{Analytic expressions for $B_{i}$ and $C_{i,j}$}\label{app:MCE}
In this appendix, we first give the explicit results for $B^{LO}_{i}$ and $C^{LO}_{i,j}$, i.e., the SM contributions to the polarization and correlation coefficients at LO.
Due to CP symmetry, $\bar{B}^{LO}_{i} = B^{LO}_{i}$ and $C^{LO}_{i,j} = C^{LO}_{j,i}$.
Therefore, we only give expressions for $B^{LO}_{i}$ and for $C^{LO}_{i,j}$ with $i \leq j$. Other polarization
and correlation coefficients that are not listed here vanish at LO.
\begin{equation}
    \begin{split}
        &B^{LO}_3 = \frac{512 \sqrt{6} \pi ^2 \alpha ^2 \left(\beta ^2-1\right) \left(y^2-1\right)}{\left(\beta ^2 y^2-1\right)^2},\\
        &B^{LO}_7 = \frac{1024 \sqrt{6} \pi ^2 \alpha ^2 y \sqrt{\left(\beta ^2-1\right) \left(y^2-1\right)}}{\left(\beta ^2 y^2-1\right)^2},\\
        &B^{LO}_8 = -\frac{512 \sqrt{2} \pi ^2 \alpha ^2 \left(\beta ^2 \left(y^2-3\right)-3 y^2+1\right)}{\left(\beta ^2 y^2-1\right)^2}.
    \end{split}
\end{equation}

\begin{equation}
    \begin{split}
        &C^{LO}_{1,1} = \frac{384 \pi ^2 \alpha ^2 \left(\beta ^4 \left(y^4-2 y^2+2\right)+\beta ^2 \left(-2 y^4+4 y^2+4\right)-8 y^2+1\right)}{\left(\beta ^2 y^2-1\right)^2},\\
        &C^{LO}_{1,5} = -\frac{768 \pi ^2 \alpha ^2 y \sqrt{\left(\beta ^2-1\right) \left(y^2-1\right)} \left(\beta ^2 \left(y^2-1\right)+4\right)}{\left(\beta ^2 y^2-1\right)^2},\\
        &C^{LO}_{2,2} = -\frac{384 \pi ^2 \alpha ^2 \left(\beta ^4 \left(y^4-2 y^2+2\right)-2 \beta ^2 \left(y^4-2 y^2+2\right)+1\right)}{\left(\beta ^2 y^2-1\right)^2},\\
        &C^{LO}_{2,4} = -\frac{768 \pi ^2 \alpha ^2 \beta ^2 y \left(y^2-1\right) \sqrt{\left(\beta ^2-1\right) \left(y^2-1\right)}}{\left(\beta ^2 y^2-1\right)^2},\\
        &C^{LO}_{3,3} = \frac{384 \pi ^2 \alpha ^2 \left(\beta ^4 \left(y^4-2 y^2+2\right)-2 \beta ^2 \left(y^4-2 y^2+2\right)+2 y^4-4 y^2+3\right)}{\left(\beta ^2 y^2-1\right)^2},\\
        &C^{LO}_{3,7} = \frac{768 \pi ^2 \alpha ^2 \left(\beta ^2-2\right) y \left(y^2-1\right) \sqrt{\left(\beta ^2-1\right) \left(y^2-1\right)}}{\left(\beta ^2 y^2-1\right)^2},\\
        &C^{LO}_{3,8} = \frac{256 \sqrt{3} \pi ^2 \alpha ^2 \left(\beta ^2-1\right) \left(3 y^4-4 y^2+1\right)}{\left(\beta ^2 y^2-1\right)^2},\\
        &C^{LO}_{4,4} = \frac{384 \pi ^2 \alpha ^2 \left(\beta ^4 \left(y^4-2 y^2+2\right)-2 \beta ^2 y^2 \left(y^2-1\right)-1\right)}{\left(\beta ^2 y^2-1\right)^2},\\
        &C^{LO}_{5,5} = -\frac{384 \pi ^2 \alpha ^2 \left(\beta ^4 \left(y^4-2 y^2+2\right)-2 \beta ^2 \left(y^4-5 y^2+4\right)-8 y^2+7\right)}{\left(\beta ^2 y^2-1\right)^2},\\
        &C^{LO}_{6,6} = \frac{384 \pi ^2 \alpha ^2 \left(\beta ^4 \left(y^4-2 y^2+2\right)+2 \beta ^2 \left(y^2-1\right)-1\right)}{\left(\beta ^2 y^2-1\right)^2},\\
        &C^{LO}_{7,7} = -\frac{384 \pi^2 \alpha ^2 \left(\beta ^4 \left(y^4-2 y^2+2\right)-2 \beta ^2 \left(4 y^4-5 y^2+1\right)+8 y^4-8 y^2-1\right)}{\left(\beta ^2 y^2-1\right)^2},\\
        &C^{LO}_{7,8} = -\frac{256 \sqrt{3} \pi^2 \alpha ^2 y \sqrt{\left(\beta ^2-1\right) \left(y^2-1\right)} \left(3 \beta ^2 \left(y^2-1\right)-6 y^2+2\right)}{\left(\beta ^2 y^2-1\right)^2},\\
        &C^{LO}_{8,8} = \frac{128 \pi ^2 \alpha ^2 \left(3 \beta ^4 \left(y^4-2 y^2+2\right)+\beta ^2 \left(20 y^2-18 y^4\right)+18 y^4-12 y^2+5\right)}{\left(\beta ^2 y^2-1\right)^2}.
    \end{split}
\end{equation}

Then, we give the contributions to the polarization and correlation coefficients from AQGCs under the linear approximation. Since
no CP violation effect is introduced in Eq.~(\ref{eq:effective-lagrangian}), $\bar{B}^{NP}_i = B^{NP}_i$ and $C^{NP}_{i,j} = C^{NP}_{j,i}$. Accordingly, the corresponding expressions for
$\bar{B}^{NP}_i$ and for $C_{i,j}^{NP}$ with $i > j$ are not given. All the other coefficients that are not given vanish at this order.

\begin{equation}
    \frac{\Lambda^2}{s}B^{NP}_{6} = a_c\frac{16 \pi ^2 \alpha ^2 \beta ^2 y \sqrt{6-6 y^2}}{\sqrt{1-\beta ^2} \left(\beta ^2 y^2-1\right)}.
    \label{eq:AnomalousPolarization}
\end{equation}

\begin{equation}
    \begin{split}
        &\frac{\Lambda^2}{s}C^{NP}_{1,2} = a_c\frac{12 \pi ^2 \alpha ^2 \beta ^2 \left(y^4-1\right)}{\beta ^2 y^2-1},\\
        &\frac{\Lambda^2}{s}C^{NP}_{1,4} = -a_c\frac{12 \pi ^2 \alpha ^2 \beta ^2 y \left(y^2+3\right) \sqrt{\frac{y^2-1}{\beta ^2-1}}}{\beta ^2 y^2-1},\\
        &\frac{\Lambda^2}{s}C^{NP}_{2,5} = -a_c\frac{12 \pi ^2 \alpha ^2 \beta ^2 y \left(1-y^2\right)^{3/2}}{\sqrt{1-\beta ^2} \left(\beta ^2 y^2-1\right)},\\
        &\frac{\Lambda^2}{s}C^{NP}_{3,6} = a_c\frac{12 \pi ^2 \alpha ^2 \beta ^2 y \left(1-y^2\right)^{3/2}}{\sqrt{1-\beta ^2} \left(\beta ^2 y^2-1\right)},\\
        &\frac{\Lambda^2}{s}C^{NP}_{4,5} = -a_0\frac{192 \pi ^2 \alpha ^2 \beta ^2}{\left(\beta ^2-1\right) (\beta  y-1) (\beta  y+1)}
        -a_c\frac{12 \pi ^2 \alpha ^2 \beta ^2 \left(-\beta ^2 \left(y^2-1\right)+y^4+y^2+2\right)}{\left(\beta ^2-1\right) (\beta  y-1) (\beta  y+1)},\\
        &\frac{\Lambda^2}{s}C^{NP}_{6,7} = -a_0\frac{192 \pi ^2 \alpha ^2 \beta ^2}{\left(\beta ^2-1\right) (\beta  y-1) (\beta  y+1)}
        -a_c\frac{12 \pi ^2 \alpha ^2 \beta ^2 \left(\left(\beta ^2-2\right) y^4-\left(\beta ^2-1\right) y^2+5\right)}{\left(\beta ^2-1\right) (\beta  y-1) (\beta  y+1)},\\
        &\frac{\Lambda^2}{s}C^{NP}_{6,8} = a_c\frac{4 \pi ^2 \alpha ^2 \beta ^2 y \sqrt{3-3 y^2} \left(3 y^2+1\right)}{\sqrt{1-\beta ^2} \left(\beta ^2 y^2-1\right)}.
    \end{split}
    \label{eq:AnomalousCorrelations}
\end{equation}

Finally, we note the relations
\begin{equation}
    C^{NP}_{2,5} + C^{NP}_{3,6} = 0,\quad \text{and}\quad
    \frac{\Lambda^2}{s}\left(C^{NP}_{4,5} - C^{NP}_{6,7}\right) = a_c\frac{12 \pi ^2 \alpha ^2 \beta ^2 \left(\beta ^2-3\right) \left(y^4-1\right)}{\left(\beta ^2-1\right) (\beta  y-1) (\beta  y+1)}.
\end{equation}
\end{appendix}

\bibliography{ref}

@article{Belanger:1992qi,
    author = "Belanger, G. and Boudjema, F.",
    title = "{gamma gamma ---{\ensuremath{>}} W+ W- and gamma gamma ---{\ensuremath{>}} Z Z as tests of novel quartic couplings}",
    reportNumber = "ENSLAPP-A-364-92, UDEM-LPN-TH-80",
    doi = "10.1016/0370-2693(92)91979-J",
    journal = "Phys. Lett. B",
    volume = "288",
    pages = "210--220",
    year = "1992"
}

@article{Pierzchala:2008xc,
    author = "Pierzchala, Tomasz and Piotrzkowski, Krzysztof",
    editor = "d'Enterria, D. and Klasen, M. and Piotrzkowski, K.",
    title = "{Sensitivity to anomalous quartic gauge couplings in photon-photon interactions at the LHC}",
    eprint = "0807.1121",
    archivePrefix = "arXiv",
    primaryClass = "hep-ph",
    reportNumber = "CP3-08-32",
    doi = "10.1016/j.nuclphysbps.2008.07.032",
    journal = "Nucl. Phys. B Proc. Suppl.",
    volume = "179-180",
    pages = "257--264",
    year = "2008"
}

@article{CMS:2013hdf,
    author = "Chatrchyan, Serguei and others",
    collaboration = "CMS",
    title = "{Study of Exclusive Two-Photon Production of $W^+W^-$ in $pp$ Collisions at $\sqrt{s} = 7$ TeV and Constraints on Anomalous Quartic Gauge Couplings}",
    eprint = "1305.5596",
    archivePrefix = "arXiv",
    primaryClass = "hep-ex",
    reportNumber = "CMS-FSQ-12-010, CERN-PH-EP-2013-084",
    doi = "10.1007/JHEP07(2013)116",
    journal = "JHEP",
    volume = "07",
    pages = "116",
    year = "2013"
}

@article{CMS:2016rtz,
    author = "Khachatryan, Vardan and others",
    collaboration = "CMS",
    title = "{Evidence for exclusive $\gamma\gamma \to W^+ W^-$ production and constraints on anomalous quartic gauge couplings in $pp$ collisions at $ \sqrt{s}=7 $ and 8 TeV}",
    eprint = "1604.04464",
    archivePrefix = "arXiv",
    primaryClass = "hep-ex",
    reportNumber = "CMS-FSQ-13-008, CERN-EP-2016-073",
    doi = "10.1007/JHEP08(2016)119",
    journal = "JHEP",
    volume = "08",
    pages = "119",
    year = "2016"
}

@article{Chapon:2009hh,
    author = "Chapon, E. and Royon, C. and Kepka, O.",
    title = "{Anomalous quartic W W gamma gamma, Z Z gamma gamma, and trilinear WW gamma couplings in two-photon processes at high luminosity at the LHC}",
    eprint = "0912.5161",
    archivePrefix = "arXiv",
    primaryClass = "hep-ph",
    doi = "10.1103/PhysRevD.81.074003",
    journal = "Phys. Rev. D",
    volume = "81",
    pages = "074003",
    year = "2010"
}

@article{Shao:2025bma,
    author = "Shao, Hua-Sheng and Simon, Lukas",
    title = "{Automated next-to-leading order QCD and electroweak predictions of photon-photon processes in ultraperipheral collisions}",
    eprint = "2504.10104",
    archivePrefix = "arXiv",
    primaryClass = "hep-ph",
    doi = "10.1007/JHEP07(2025)020",
    journal = "JHEP",
    volume = "07",
    pages = "020",
    year = "2025"
}

@article{Ashby-Pickering:2022umy,
    author = "Ashby-Pickering, Rachel and Barr, Alan J. and Wierzchucka, Agnieszka",
    title = "{Quantum state tomography, entanglement detection and Bell violation prospects in weak decays of massive particles}",
    eprint = "2209.13990",
    archivePrefix = "arXiv",
    primaryClass = "quant-ph",
    doi = "10.1007/JHEP05(2023)020",
    journal = "JHEP",
    volume = "05",
    pages = "020",
    year = "2023"
}

@article{Denner:1991kt,
    author = "Denner, Ansgar",
    title = "{Techniques for calculation of electroweak radiative corrections at the one loop level and results for W physics at LEP-200}",
    eprint = "0709.1075",
    archivePrefix = "arXiv",
    primaryClass = "hep-ph",
    reportNumber = "PRINT-91-0349 (WURZBURG)",
    doi = "10.1002/prop.2190410402",
    journal = "Fortsch. Phys.",
    volume = "41",
    pages = "307--420",
    year = "1993"
}

@article{Denner:2019vbn,
    author = "Denner, Ansgar and Dittmaier, Stefan",
    title = "{Electroweak Radiative Corrections for Collider Physics}",
    eprint = "1912.06823",
    archivePrefix = "arXiv",
    primaryClass = "hep-ph",
    reportNumber = "FR-PHENO-019",
    doi = "10.1016/j.physrep.2020.04.001",
    journal = "Phys. Rept.",
    volume = "864",
    pages = "1--163",
    year = "2020"
}

@article{Denner:2023ehn,
    author = "Denner, Ansgar and Haitz, Christoph and Pelliccioli, Giovanni",
    title = "{NLO EW corrections to polarised W+W{\ensuremath{-}} production and decay at the LHC}",
    eprint = "2311.16031",
    archivePrefix = "arXiv",
    primaryClass = "hep-ph",
    reportNumber = "COMETA-2023-01, MPP-2023-268",
    doi = "10.1016/j.physletb.2024.138539",
    journal = "Phys. Lett. B",
    volume = "850",
    pages = "138539",
    year = "2024"
}

@article{Frixione:1993yw,
    author = "Frixione, Stefano and Mangano, Michelangelo L. and Nason, Paolo and Ridolfi, Giovanni",
    title = "{Improving the Weizsacker-Williams approximation in electron - proton collisions}",
    eprint = "hep-ph/9310350",
    archivePrefix = "arXiv",
    reportNumber = "CERN-TH-7032-93, GEF-TH-18-93",
    doi = "10.1016/0370-2693(93)90823-Z",
    journal = "Phys. Lett. B",
    volume = "319",
    pages = "339--345",
    year = "1993"
}

@article{Baltz:2007kq,
    author = "Baltz, A. J. and others",
    title = "{The Physics of Ultraperipheral Collisions at the LHC}",
    eprint = "0706.3356",
    archivePrefix = "arXiv",
    primaryClass = "nucl-ex",
    doi = "10.1016/j.physrep.2007.12.001",
    journal = "Phys. Rept.",
    volume = "458",
    pages = "1--171",
    year = "2008"
}

@article{Denner:2005nn,
    author = "Denner, Ansgar and Dittmaier, S.",
    title = "{Reduction schemes for one-loop tensor integrals}",
    eprint = "hep-ph/0509141",
    archivePrefix = "arXiv",
    reportNumber = "MPP-2005-84, PSI-PR-05-08",
    doi = "10.1016/j.nuclphysb.2005.11.007",
    journal = "Nucl. Phys. B",
    volume = "734",
    pages = "62--115",
    year = "2006"
}

@article{Trzebinski:2023lzg,
    author = "Trzebi{\'n}ski, Maciej",
    collaboration = "ATLAS Forward Detectors",
    title = "{Overview of ATLAS forward proton detectors for LHC Run 3 and plans for the HL-LHC}",
    eprint = "2310.11783",
    archivePrefix = "arXiv",
    primaryClass = "hep-ex",
    reportNumber = "ATL-FWD-PROC-2022-001",
    doi = "10.1016/j.nima.2023.168187",
    journal = "Nucl. Instrum. Meth. A",
    volume = "1050",
    pages = "168187",
    year = "2023"
}

@article{Bossini:2020ycc,
    author = "Bossini, E.",
    editor = "Nessi, Marzio",
    collaboration = "CMS, TOTEM",
    title = "{The CMS Precision Proton Spectrometer timing system: performance in Run 2, future upgrades and sensor radiation hardness studies}",
    eprint = "2004.11068",
    archivePrefix = "arXiv",
    primaryClass = "physics.ins-det",
    reportNumber = "CMS-CR-2019-294",
    doi = "10.1088/1748-0221/15/05/C05054",
    journal = "JINST",
    volume = "15",
    number = "05",
    pages = "C05054",
    year = "2020"
}

@inproceedings{Denner:1996wm,
    author = "Denner, Ansgar and Dittmaier, S. and Schuster, R.",
    title = "{Electroweak radiative corrections to gamma gamma ---{\ensuremath{>}} W+ W-}",
    booktitle = "{Physics with e+ e- Linear Colliders (The European Working Groups 4 Feb - 1 Sep 1995: Session 1 (Session 2: 2-3 Jun 1995 in Assergi, Italy: Session 3: 30 Aug - 1 Sep 1995 in Hamburg, Germany)}",
    eprint = "hep-ph/9601355",
    archivePrefix = "arXiv",
    reportNumber = "BI-TP-96-03, WUE-ITP-96-001",
    pages = "233--240",
    month = "1",
    year = "1996"
}

@misc{Eboli:2026mbq,
    author = "{\'E}boli, Oscar J. P. and Rahaman, Rafiqul and Subba, Amir",
    title = "{Probing anomalous quartic gauge couplings in same-sign $W$ boson scattering with polarization and spin correlation}",
    eprint = "2606.06436",
    archivePrefix = "arXiv",
    primaryClass = "hep-ph",
    month = "6",
    year = "2026"
}

@article{Jikia:1996uu,
    author = "Jikia, G.",
    title = "{Electroweak O (alpha) corrections to W+ W- pair production in polarized gamma gamma collisions}",
    eprint = "hep-ph/9612380",
    archivePrefix = "arXiv",
    reportNumber = "FREIBURG-THEP-96-23",
    doi = "10.1016/S0550-3213(97)00170-3",
    journal = "Nucl. Phys. B",
    volume = "494",
    pages = "19--40",
    year = "1997"
}

@article{Denner:1995jv,
    author = "Denner, Ansgar and Dittmaier, S. and Schuster, R.",
    title = "{Radiative corrections to $\gamma \gamma \to W^{+} W^{-}$ in the electroweak standard model}",
    eprint = "hep-ph/9503442",
    archivePrefix = "arXiv",
    reportNumber = "WUE-ITP-95-01, BI-TP-95-04",
    doi = "10.1016/0550-3213(95)00344-R",
    journal = "Nucl. Phys. B",
    volume = "452",
    pages = "80--108",
    year = "1995"
}

@article{D0:2013rce,
    author = "Abazov, Victor Mukhamedovich and others",
    collaboration = "D0",
    title = "{Search for Anomalous Quartic $WW{\gamma}{\gamma}$ Couplings in Dielectron and Missing Energy Final States in $p\bar{p}$ Collisions at $\sqrt{s}$ = 1.96 TeV}",
    eprint = "1305.1258",
    archivePrefix = "arXiv",
    primaryClass = "hep-ex",
    reportNumber = "FERMILAB-PUB-13-133-E",
    doi = "10.1103/PhysRevD.88.012005",
    journal = "Phys. Rev. D",
    volume = "88",
    pages = "012005",
    year = "2013"
}

@article{Lysenko:2026hql,
    author = "Lysenko, Viktoriia",
    collaboration = "ATLAS Forward Detectors",
    title = "{Overview of ATLAS forward proton detectors: status, performance and new physics results}",
    doi = "10.22323/1.485.0529",
    journal = "PoS",
    volume = "EPS-HEP2025",
    pages = "529",
    year = "2026"
}

@article{ATLAS:2016lse,
    author = "Aaboud, Morad and others",
    collaboration = "ATLAS",
    title = "{Measurement of exclusive $\gamma\gamma\rightarrow W^+W^-$ production and search for exclusive Higgs boson production in $pp$ collisions at $\sqrt{s} = 8$ TeV using the ATLAS detector}",
    eprint = "1607.03745",
    archivePrefix = "arXiv",
    primaryClass = "hep-ex",
    reportNumber = "CERN-EP-2016-123",
    doi = "10.1103/PhysRevD.94.032011",
    journal = "Phys. Rev. D",
    volume = "94",
    number = "3",
    pages = "032011",
    year = "2016"
}

@article{ATLAS:2016snd,
    author = "Aaboud, Morad and others",
    collaboration = "ATLAS",
    title = "{Measurement of $W^{\pm}W^{\pm}$ vector-boson scattering and limits on anomalous quartic gauge couplings with the ATLAS detector}",
    eprint = "1611.02428",
    archivePrefix = "arXiv",
    primaryClass = "hep-ex",
    reportNumber = "CERN-EP-2016-167",
    doi = "10.1103/PhysRevD.96.012007",
    journal = "Phys. Rev. D",
    volume = "96",
    number = "1",
    pages = "012007",
    year = "2017"
}

@article{ATLAS:2020iwi,
    author = "Aad, Georges and others",
    collaboration = "ATLAS",
    title = "{Observation of photon-induced $W^+W^-$ production in $pp$ collisions at $\sqrt{s}=13$ TeV using the ATLAS detector}",
    eprint = "2010.04019",
    archivePrefix = "arXiv",
    primaryClass = "hep-ex",
    reportNumber = "CERN-EP-2020-165",
    doi = "10.1016/j.physletb.2021.136190",
    journal = "Phys. Lett. B",
    volume = "816",
    pages = "136190",
    year = "2021"
}

@article{CMS:2022dmc,
    author = "Tumasyan, Armen and others",
    collaboration = "CMS, TOTEM",
    title = "{Search for high-mass exclusive $\gamma\gamma\to WW$ and $\gamma\gamma\to ZZ$ production in proton-proton collisions at $\sqrt{s}$ = 13 TeV}",
    eprint = "2211.16320",
    archivePrefix = "arXiv",
    primaryClass = "hep-ex",
    reportNumber = "CMS-SMP-21-014, TOTEM-2022-002, CERN-EP-2022-177",
    doi = "10.1007/JHEP07(2023)229",
    journal = "JHEP",
    volume = "07",
    pages = "229",
    year = "2023"
}

@article{CMS:2026job,
    author = "Hayrapetyan, Aram and others",
    collaboration = "CMS",
    title = "{Measurement and effective field theory interpretation of the photon-fusion production cross section of a pair of W bosons in proton-proton collisions at $ \sqrt{s}=13 $ TeV}",
    eprint = "2601.21574",
    archivePrefix = "arXiv",
    primaryClass = "hep-ex",
    reportNumber = "CMS-SMP-24-019, CERN-EP-2025-273",
    doi = "10.1007/JHEP06(2026)187",
    journal = "JHEP",
    volume = "06",
    pages = "187",
    year = "2026"
}

@article{Hahn:2000kx,
    author = "Hahn, Thomas",
    title = "{Generating Feynman diagrams and amplitudes with FeynArts 3}",
    eprint = "hep-ph/0012260",
    archivePrefix = "arXiv",
    reportNumber = "KA-TP-23-2000",
    doi = "10.1016/S0010-4655(01)00290-9",
    journal = "Comput. Phys. Commun.",
    volume = "140",
    pages = "418--431",
    year = "2001"
}

@article{Shtabovenko:2023idz,
    author = "Shtabovenko, Vladyslav and Mertig, Rolf and Orellana, Frederik",
    title = "{FeynCalc 10: Do multiloop integrals dream of computer codes?}",
    eprint = "2312.14089",
    archivePrefix = "arXiv",
    primaryClass = "hep-ph",
    reportNumber = "P3H-23-089, TTP23-056, SI-HEP-2023-27",
    doi = "10.1016/j.cpc.2024.109357",
    journal = "Comput. Phys. Commun.",
    volume = "306",
    pages = "109357",
    year = "2025"
}

@article{Hahn:1998yk,
    author = "Hahn, T. and Perez-Victoria, M.",
    title = "{Automatized one loop calculations in four-dimensions and D-dimensions}",
    eprint = "hep-ph/9807565",
    archivePrefix = "arXiv",
    reportNumber = "UG-FT-87-98, KA-TP-7-1998",
    doi = "10.1016/S0010-4655(98)00173-8",
    journal = "Comput. Phys. Commun.",
    volume = "118",
    pages = "153--165",
    year = "1999"
}

@article{Hahn:2004fe,
    author = "Hahn, T.",
    title = "{CUBA: A Library for multidimensional numerical integration}",
    eprint = "hep-ph/0404043",
    archivePrefix = "arXiv",
    reportNumber = "MPP-2004-40",
    doi = "10.1016/j.cpc.2005.01.010",
    journal = "Comput. Phys. Commun.",
    volume = "168",
    pages = "78--95",
    year = "2005"
}

@article{Passarino:1978jh,
    author = "Passarino, G. and Veltman, M. J. G.",
    title = "{One Loop Corrections for e+ e- Annihilation Into mu+ mu- in the Weinberg Model}",
    reportNumber = "Print-79-0284 (UTRECHT)",
    doi = "10.1016/0550-3213(79)90234-7",
    journal = "Nucl. Phys. B",
    volume = "160",
    pages = "151--207",
    year = "1979"
}

@article{ParticleDataGroup:2026mpi,
    author = "Takahashi, F. and others",
    collaboration = "Particle Data Group",
    title = "{Review of Particle Physics}",
    doi = "10.1142/s0217751x26300115",
    journal = "Int. J. Mod. Phys. A",
    volume = "41",
    number = "22",
    pages = "2630011",
    year = "2026"
}

@article{Denner:2020bcz,
    author = "Denner, Ansgar and Pelliccioli, Giovanni",
    title = "{Polarized electroweak bosons in ${\bf \text{W}^+\text{W}^-}$ production at the LHC including NLO QCD effects}",
    eprint = "2006.14867",
    archivePrefix = "arXiv",
    primaryClass = "hep-ph",
    doi = "10.1007/JHEP09(2020)164",
    journal = "JHEP",
    volume = "09",
    pages = "164",
    year = "2020"
}

@article{Bredenstein:2005zk,
    author = "Bredenstein, A. and Dittmaier, S. and Roth, M.",
    title = "{Four-fermion production at gamma gamma colliders. 2. Radiative corrections in double-pole approximation}",
    eprint = "hep-ph/0506005",
    archivePrefix = "arXiv",
    reportNumber = "MPP-2005-24",
    doi = "10.1140/epjc/s2005-02343-5",
    journal = "Eur. Phys. J. C",
    volume = "44",
    pages = "27--49",
    year = "2005"
}

\end{document}